\documentclass[aps,prd,preprint,tightenlines,superscriptaddress,
amsfonts,amssymb,amsmath,showpacs,nofootinbib]{revtex4-1}
\usepackage[polish,english]{babel}
\usepackage[utf8]{inputenc}
\usepackage[T1]{fontenc}
\usepackage{latexsym}
\usepackage{graphicx}
\usepackage{geometry}
\usepackage{epstopdf}
\usepackage{color}
\DeclareMathOperator*{\SumInt}{%
\mathchoice%
  {\ooalign{$\displaystyle\sum$\cr\hidewidth$\displaystyle\int$\hidewidth\cr}}
  {\ooalign{\raisebox{.14\height}{\scalebox{.7}{$\textstyle\sum$}}\cr\hidewidth$\textstyle\int$\hidewidth\cr}}
  {\ooalign{\raisebox{.2\height}{\scalebox{.6}{$\scriptstyle\sum$}}\cr$\scriptstyle\int$\cr}}
  {\ooalign{\raisebox{.2\height}{\scalebox{.6}{$\scriptstyle\sum$}}\cr$\scriptstyle\int$\cr}}
}

\usepackage{xcolor}
\usepackage{pdfsync}
\usepackage{fancyhdr}
\usepackage{makeidx}
\usepackage[breaklinks,colorlinks,backref]{hyperref}
\hypersetup{
    colorlinks, %set true if you want colored links
    linktoc=all, %set to all if you want both sections and subsections linked
    linkcolor=red,  %choose some color if you want links to stand out
  citecolor=hyptxt,
  urlcolor=blue}
  \hypersetup{
  citebordercolor=,
  filebordercolor=green,
  linkbordercolor=blue
}
\definecolor{hyptxt}{rgb}{0.7, 0.4, 0.9}
\usepackage{bibtopic}

\newcommand{\dR}{\mathbb R}

\newcommand{\id}{\mathbb I}

\newcommand{\be}{\begin{equation}}
\newcommand{\ee}{\end{equation}}
\newcommand{\R}{\mathbb R}

\newcommand{\ket}[1]{|\kern.3ex#1\kern.3ex\rangle}
\newcommand{\bra}[1]{\langle\kern.3ex #1 \kern.3ex|}
\newcommand{\scalar}[2]{\langle\kern.3ex #1 \kern.3ex|\kern.3ex#2\kern.3ex\rangle}
\newcommand{\norm}[1]{\|\kern.3ex#1\kern.3ex \|}

\newcommand{\UnitOp}{\hat{\mathbb{I}}} % operator unit with hat
\newcommand{\Group}[1]{\mathrm{#1}} % group assignment
\newcommand{\Ket}[1]{\vert #1 \rangle} % ket
\newcommand{\BraKet}[2]{\langle #1 \vert #2 \rangle} % bra(c)ket
\newcommand{\Komentarz}[1]{}  % comment

\begin{document}
%tightenlines - single space manuscript
%eqsecnum - number equations by section

\title{Hunting for gravitational quantum spikes}

\author{Andrzej G\'{o}\'{z}d\'{z}}
\email{andrzej.gozdz@umcs.lublin.pl}
\affiliation{Institute of Physics, Maria Curie-Sk{\l}odowska
University, pl.  Marii Curie-Sk{\l}odowskiej 1, 20-031 Lublin, Poland}

\author{W{\l}odzimierz Piechocki} \email{wlodzimierz.piechocki@ncbj.gov.pl}
\affiliation{Department of Fundamental Research, National Centre for Nuclear
  Research, Pasteura 7, 02-093 Warsaw, Poland  }

\author{Grzegorz Plewa} \email{grzegorz.plewa@ncbj.gov.pl}
\affiliation{Department of Fundamental Research, National Centre for Nuclear
  Research, Pasteura 7, 02-093 Warsaw, Poland  }

\author{Tomasz Trze\'{s}niewski} \email{t.trzesniewski@uj.edu.pl}
\affiliation{Institute of  Theoretical  Physics, Jagiellonian University, {\L}ojasiewicza 11, 30-348 Krak\'{o}w, Poland}

\date{\today}

\begin{abstract}  We present the result of our examination of quantum
    structures called quantum spikes. The classical spikes, that are
    known in gravitational systems, occur in the evolution of the inhomogeneous
    spacetimes.  Different kind of spikes, which we name strange spikes, can be
    seen in the dynamics of the homogeneous sector of the
    Belinski-Khalatnikov-Lifshitz scenario. They can be made visible if the
    so-called inhomogeneous initial data are used. The question to be explored
    is whether the strange spikes may survive quantization. The answer is in the
    affirmative. However, this is rather a subtle effect that needs further
    examination using sophisticated analytical and numerical tools. The spikes
    seem to be of fundamental importance, both at classical and quantum levels,
    as they may serve as seeds of real structures in the universe.

\end{abstract}

%\pacs{abc}

\maketitle

\tableofcontents

%%%%%%%%%%%%%%%%%%%%%%%%%%%%%%%%%%%%%%%%%%%%%%%%%%%%%%%%%%%%%%%%%%%%%%%%%%
\section{Introduction}

The dynamics underlying the Belinskii-Khalatnikov-Lifshitz (BKL) scenario, which
concerns a generic gravitational singularity \cite{BKL22,BKL33}, can be
described by the nonlinear coupled system of ODEs for the three effective
directional scale factors (see Part~I of \cite{BKL44}). This dynamics has been
recently quantized \cite{AWG,AW}. The quantum BKL scenario predicts that a
gravitational singularity can be avoided by a quantum bounce, occurring in the
unitary evolution of a given gravitational system.

A different approach to solve the problem of a singularity in the BKL scenario
has been proposed by Ashtekar et al. \cite{Ashtekar:2011ck}. It can be used,
after the successful quantization, to tackle a generic gravitational singularity
as well. Furthermore, even if one restricts to its homogeneous sector, the model
can be explored from the perspective of another interesting issue, which is the
emergence of gravitational structures known as spikes. The aim of this paper is
to uncover such structures at classical level and to investigate if they can
survive the quantization. The spikes that we name here the ``strange spikes''
are different from the (transient or permanent) spikes observed in the dynamics
of inhomogeneous spacetimes (see \cite{sp1,sp2,sp3,sp4,sp5,sp6,sp7,sp8,sp9} and
references therein). The latter have well understood properties, whereas our
spikes have been discovered and preliminarily examined in the context of quantum
physics only recently, in \cite{Czuchry:2016rlo}. Results of the latter paper
suggest that quantum (strange) spikes do not exist. However, the issue of time
has not been treated satisfactorily due to the fact that \cite{Czuchry:2016rlo}
deals mainly with the vacuum case. In the present paper, we couple the system to
a massless scalar field, so that it can be used as a reference clock at both the
classical and quantum levels. Moreover, \cite{Czuchry:2016rlo} has included only
simplified analyses of quantum observables of a spike. Our paper feels this gap
as well.

Let us stress that we do not address the issue of possible resolution of a
generic gravitational singularity, which is predicted by the BKL conjecture and
follows from the quantization of the {\it full} classical dynamics. Instead, we
examine the possibility of formation of spikes resulting from the nonlinearity
of the dynamics that is specific to its {\it homogeneous} sector, as defined in
\cite{Ashtekar:2011ck} by Eqs.~(5.7)--(5.11). The latter is the total dynamics
that is intended to be the subject of our paper.

The classical and quantum spikes that we examine are subtle structures, which
appear in a rather complicated dynamics at both the classical and quantum
levels. They are of fundamental importance as they may serve as seeds of
macro-structures in Universe (like, e.g., filaments built from superclusters of
galaxies) in the former case, and quantum fluctuations (which may underly, e.g.,
the creation of primordial black holes) in the latter case.

On the other hand, such structures are specific to any nonlinear coupled system
of ODEs in which one considers the mapping of a smooth curve of the initial data
into another curve via the propagation by the same amount of time at each point
of the initial data. It may happen that the initial curve evolves into an
intriguing structure that we call a strange spike.

Spikes occur in a variety of dynamical systems. For instance, in
  the context of dynamics of a forced pendulum with damping. The damped driven
(forced) pendulum models have applications in mathematical biology (see
\cite{ThH,neuro} and references therein).
%The present paper is devoted to the
%examination of the possible structure formation, both at classical and quantum
%levels, in the context of gravitational physics. {\bf [a repetition - TT]}
In general, the name ``spike'' is used in literature in very different physical
and mathematical contexts. Actually, in most cases when the function describing
a given phenomenon has a jump in some region, the latter is called a spike. An
interesting problem is the existence of the quantum spikes.
The paper by A.~Tilloy et al. \cite{Tilloy2015} contains a particular example of
them. The authors define quantum spikes as a certain kind of quantum
fluctuations in the system that are able to jump between different states and
can be described by a set of stochastic equations. This type of spikes, which
probably can be observed in any quantum system satisfying required conditions,
is different from quantum gravitational spikes presented in our paper.

As we already mentioned, an important aspect of this work is the problem of
time, which occurs when quantization is applied to variables describing the
dynamical spacetime geometry. At the classical level, we use the well known
formalism of parameterizing time via a scalar field, which acts as a reference
clock. We propose to choose the corresponding reference clock at the quantum
level, which requires introducing some special mathematical structure. This
construction leads to a specific formalism describing the quantum dynamics.  All
details are presented in Sec.~IIIB.

The paper is organized as follows: In Sec.~II, we define the phase space
variables satisfying the affine Lie algebra, Hamilton dynamics parameterized by
a massless scalar field, and derive the classical spikes. Sec.~III deals with
the quantization. We specify the representation of the Lie algebra and introduce
the quantum evolution parameter. We represent the quantum evolution in terms of
two eigenequations and find numerical solutions to these equations.  The quantum
dynamical constraint is imposed. In Sec.~IV we derive the quantum
spikes. Conclusions are presented in Sec.~V.

%%%%%%%%%%%%%%%%%%%%%%%%%%%%%%%%%%%%%%%%%%%%%%%%%%%%%%%%%%%%%%%%%%%%%%%%%%
\section{Classical level}

%%%%%%%%%%%%%%%%%%%%%%%%%%%%%%%%%%%%%%%%%%%%%%%%%%%%%%%%%%%%%%%%%%%%%%%%%
\subsection{Phase space}

The kinematical phase space of the homogeneous sector of the gravitational field
in the Hamiltonian formalism \cite{Ashtekar:2011ck} can be parameterized by the
variables $C_I$ and $P_I$, with $I = 1,2,3$.  Each of these variables is defined
as the integral of a (homogeneous) field over the spatial hypersurface. For
details, see in particular section V of \cite{Ashtekar:2011ck}. We also
introduce a massless scalar field, described by the variables $\phi$ and $\pi$
(also defined as integrated over space), where $\pi$ is the conjugate momentum.
Poisson brackets for the total system read \cite{Ashtekar:2011ck}
\begin{align}\label{a1}
 \{P_I,P_J\}= 0 = \{C_I,C_J\}\,, \qquad \{P_I,C_J\}
= 2\delta^I_J C_I\,, \qquad \{\phi,\pi\} = 1\,.
\end{align}
To connect with the notation that is more common for affine algebras, we perform
the partial redefinition of variables $(C_I,P_J) =: (C_I,-2D_J)$, which leads to
the affine Poisson brackets \cite{Czuchry:2016rlo}
\begin{align}\label{b1}
 \{D_I,D_J\} = 0 = \{C_I,C_J\}\,, \qquad \{C_J,D_I\} = \delta^I_J C_I\,.
\end{align}
An algebra with such brackets is called an affine Lie algebra.

The dynamics of the system is specified by the equations\footnote{We have the
extra factor $2$ in  Eqs. \!\eqref{b3} and \eqref{b4} that is missing in
the corresponding equations of \cite{Czuchry:2016rlo}.}

\begin{align}
% \nonumber to remove numbering (before each equation
 \dot{D}_I &= -C_I (C - 2 C_I)\,, \label{b3}\\
 \dot{C}_I &= 4 C_I (D - 2 D_I)\,, \label{b4}\\
 \dot\pi &= 0\,, \label{b5}\\
 \dot\phi &= \kappa\pi\,, \label{b6}
\end{align}
where $D = D_1 + D_2 + D_3$ and $C= C_1 + C_2 + C_3$. There is no summation
$\sum_I$ in the rhs of \eqref{b3} and \eqref{b4}. The solutions to
\eqref{b3}--\eqref{b6} have to satisfy the Hamiltonian constraint
\begin{align}\label{b7}
 H = \frac{1}{2}\, C^2 - \sum_I C_I^2
+ 4 \left(\frac{1}{2}\, D^2 - \sum_I D_I^2\right)
+ \frac{\kappa}{2}\, \pi^2 = 0\,,
\end{align}
where $\kappa=\pm 1$ defines two possible dynamics (two different signatures of
the corresponding bilinear forms) with respect to the field $\phi$. Unlike the
traditional momentum, which serves to translate the canonical coordinate $C_I$,
the variable $D_I$ serves to dilate $C_I$.

The set of equations \eqref{b3}--\eqref{b7} incorporates the dynamics of
all Bianchi-type A models. It presents a coupled system of nonlinear equations
that has not been solved in the general case analytically yet. To get some
insight into the local geometry of the space of solutions to these equations,
we apply the dynamical systems method \cite{Per,Wig}.

It is easy to see that the space $S$ of the critical points of the dynamics,
defined by the vanishing of the right hand sides of \eqref{b3}--\eqref{b6} and
satisfying the constraint \eqref{b7}, reads
\begin{equation}\label{cr1}
  S = \{(C_1,C_2,C_3, D_1,D_2,D_3,\pi, \phi)\in \R^8~|~(C_I = 0 =\pi )\wedge(D^2 = 2 \sum_I D_I^2)\} \, ,
\end{equation}
where $I=1,2,3$ and $D=D_1+D_2+D_3$.

The Jacobian of the system \eqref{b3}--\eqref{b6} is easily found to be
\begin{equation}\label{crJ}
 J=\begin{pmatrix}
 J_{11} & -C_1 & -C_1& 0& 0 & 0 & 0& 0 \\
 -C_2 & J_{22} & -C_2& 0& 0 & 0& 0& 0 \\
 -C_3 & -C_3 & J_{33}& 0& 0 & 0& 0 & 0 \\
 J_{41} & 0 & 0& -4C_1& 4C_1 & 4C_1 & 0 & 0 \\
 0 & J_{52} & 0& 4C_2& -4C_2 & 4C_2 & 0 & 0 \\
 0 & 0 & J_{63}& 4C_3& 4C_3 & -4C_3& 0 &0 \\
 0 & 0 & 0& 0 & 0 & 0& 0 & 0 \\
 0 & 0 & 0& 0 & 0 & 0& 0 & \kappa \\
 \end{pmatrix} \, ,
\end{equation}
where
\[J_{11} = 2C_1 - C_2 - C_3,~~J_{22}= -C_1+2C_2-C_3,~~J_{33}=-C_1-C_2 +2C_3 ,\]
and where
\[ J_{41}=4(-D_1+D_2+D_3),~~J_{52}=4(D_1-D_2+D_3),~~J_{63}= 4(D_1+D_2-D_3)  .\]

The Jacobian evaluated at any point of the set \eqref{cr1} is the following
matrix
\begin{equation}\label{crJS}
 J_S=\begin{pmatrix}
 0 & 0 & 0& 0& 0 & 0 & 0& 0 \\
 0 & 0 & 0& 0& 0 & 0& 0& 0 \\
 0 & 0 & 0& 0& 0 & 0& 0 & 0 \\
 J_{41} & 0 & 0& 0& 0 & 0 & 0 & 0 \\
 0 & J_{52} & 0& 0& 0 & 0 & 0 & 0 \\
 0 & 0 & J_{63}& 0& 0 & 0& 0 &0 \\
 0 & 0 & 0& 0 & 0 & 0& 0 & 0 \\
 0 & 0 & 0& 0 & 0 & 0& 0 & \kappa \\
 \end{pmatrix} \, .
\end{equation}
Thus, the characteristic polynomial associated with $J_S$ reads
\begin{equation}\label{cr2}
  P (\lambda) = (-\lambda)^7 (\kappa - \lambda) \, ,
\end{equation}
so that the eigenvalues are the following:
\begin{equation}\label{cr2}
 (0,0,0,0,0,0,0,\kappa) \, .
\end{equation}
Since the real parts of all, but one, eigenvalues of the Jacobian $J_S$ are
equal to zero, the fixed points defined by Eq. \!\eqref{cr1} are
nonhyperbolic\footnote{A critical point is called a hyperbolic fixed point if
  all eigenvalues of the Jacobian matrix of the linearized equations at this
  point have nonzero real parts. Otherwise, it is called a nonhyperbolic fixed
  point \cite{Per,Wig}.}. Thus, getting insight into the structure of the space
of solutions to the dynamics near such points require an examination of the
exact form of the dynamics.  The information obtained from linearized set of
equations is unable to reveal the nature of dynamics in the neighborhood of such
fixed points.

In the next subsection we present explicit, but approximate solution to our
dynamics characterising the strange spike.  It is obtained by solving the
dynamics with the so-called inhomogeneous initial data\footnote{This definition
and example explaining the idea of the inhomogeneous initial data are due to
David Sloan.}.  The latter means that the initial data is not just a set of 3
$C$s and 3 $D$s per point in phase space, but the related data on some curve in
this space. For instance, let us choose $(C_1, C_2, C_3) := (\tilde{x}, 0.8,
0.4)$ and $(D_1, D_2, D_3) := (f(\tilde{x}), 2, 7)$, where $f(\tilde{x})$ is the
value that solves the Hamiltonian constraint \eqref{b7} for $C_1 = \tilde{x}$.
Next, we allow $\tilde{x}$ to vary from $-0.1$ to $0.1$. Then, instead of
solving one set of equations for each point of space, we solve a whole
continuous (in practical calculations, discrete) family of them by taking the
sequence of $\tilde{x} \in (-0.1, 0.1)$. The plot of the $C$s and the $D$s as
functions of $\tilde{x}$ reveals a peculiar structure that emerges in
$\tilde{x}$ around $\tilde{x} = 0$ that we call the strange spike\footnote{We
  use $\tilde{x}$ to denote the initial data in phase space sticking to the
  notation of Sec. II of Ref.~\cite{Czuchry:2016rlo}.}

It results from Eqs. \!\eqref{b5} and \eqref{b6} that $\phi$ is a monotonic
function of time. Thus, it can be used as an evolution parameter of the
dynamics. Dividing both sides of \eqref{b3} and \eqref{b4} by $\dot\phi = \kappa
\pi$, we obtain
\begin{align}
%\nonumber to remove numbering (before each equation)
 \kappa \pi\, \frac{dD_I}{d\phi} &= -C_I (C - 2 C_I)\,, \label{ab3}\\
 \kappa \pi\, \frac{dC_I}{d\phi} &= 4C_I (D - 2 D_I)\,, \label{ab4}
\end{align}
which defines the relative dynamics with respect to the variable $\phi$.

In our case the equations of motion \eqref{ab3}--\eqref{ab4} and the
constraint \eqref{b7} are constructed from elements of the affine algebra
\eqref{b1}.  This algebra can be realized in terms of the Poisson algebra by the
adjoint action
\begin{equation}
\label{eq:AdAction}
\{X, \cdot \} Y = \{X,Y\} \  ,
\end{equation}
where $X$ and $Y$ are linear combinations of basic elements $C_J$ and $D_I$ of
affine algebra. Exponentiation of this algebra gives the operators representing
the affine group. For fixed $I$ (where $I=1,2,3$) the elements of the affine group in ``one
direction'' are represented by the following operations in the phase space
\begin{equation}
\label{eq:AffGrpElem}
g_I(\alpha_I,\beta_I)=
e^{\alpha_I \{C_I, \cdot \}} e^{-\beta_I \{D_I, \cdot \}} \ ,
\end{equation}
where $\alpha_I \in \dR $ and $\beta_I \in \dR_{+}$.

The full group is composed of all three independent transformations \\
$g(\alpha_1,\beta_1,\alpha_2,\beta_2,\alpha_3,\beta_3) =
g_1(\alpha_1,\beta_1) g_2(\alpha_2,\beta_2) g_3(\alpha_3,\beta_3)$.
This property allows, in most cases, to perform calculations for fixed $I$ and
to generalize the result to other $I$.

The action of this group on the basic elements of the affine algebra can be
summarized as
\begin{eqnarray}
\label{eq:ActAffGrpAlg}
&& g_I(\alpha_I,1) C_J= C_J, \quad
g_I(\alpha_I,1) D_J = D_J + \delta_{IJ} \alpha_J C_J \nonumber \\
&&  g_I(0,\beta_I) C_J= \delta_{IJ} \beta_J C_J, \quad
g_I(\alpha_I,1) D_J = D_J + \delta_{IJ} \alpha_J C_J.
\end{eqnarray}
The manifold of the variables $(C_I,D_I)$ splits into orbits with respect to
the affine group. The orbit of the element $(C_{I0},D_{I0})$ is defined as
\begin{equation}
\label{eq:AffGrpOrbit}
\Pi_{(C_{I0},D_{I0})}^I = \{(C_I,D_I) \colon
(C_I,D_I) = g_I(\alpha_I,\beta_I)(C_{I0},D_{I0}) \} \, .
\end{equation}
Using \eqref{eq:ActAffGrpAlg}, it is easy to check that one gets two large orbits
which we denote by $\Pi_-^I$ and $\Pi_+^I$ and continuum of orbits consisting of
single points $ \Pi_{D_I}^I$
\begin{eqnarray}
&& \Pi_{-}^I :=
\{ (C_I, D_I) \;|\; C_I \in \dR_-, D_I \in \dR \}\,, \label{b8a}\\
&& \Pi_{+}^I  :=
\{ (C_I, D_I) \;|\; C_I \in \dR_+, D_I \in \dR \}\,, \label{b8b}\\
&& \Pi_{D_I}^I := \{ (0, D_I) \}\, \text{ where } D_I \in \dR\,.  \label{b8c}
\end{eqnarray}
In fact, due to the constraint \eqref{b7} the sets of points in the orbits are
reduced to some submanifolds.

Since $C_I = 0$ is a critical point of the system \eqref{b3}--\eqref{b4}, the sign
of each $C_I$ along any dynamical trajectory is fixed by the initial
conditions. The orbits $\Pi_-^I$ and $\Pi_+^I$ carry such solutions.  Every
single-point orbit $\Pi_{D_I}^I$ separates positive $(C_I>0)$ and negative
$(C_I<0)$ parts of the kinematical trajectory. However, if the system enters
the orbit $\Pi_{D_I}^I$, it is not able to leave it. This property is very
strong. It is even fulfilled not only for infinitesimal perturbation of the
motion, but also for finite difference form of the equations of motion obtained
by Euler algorithm. Changing in \eqref{ab3}--\eqref{ab4} derivatives into finite
differences, one gets
\begin{eqnarray}
%\nonumber to remove numbering (before each equation)
&& \kappa \pi\, D_I(\phi_{n+1})
= -C_I(\phi_{n}) \left(C(\phi_{n}) - 2 C_I(\phi_{n})\right) \Delta\phi
+\kappa \pi \, D_I(\phi_{n})  \,, \label{Eab3}\\
&& \kappa \pi\, C_I(\phi_{n+1})
= 4C_I(\phi_{n}) \left( D(\phi_{n}) - 2 D_I(\phi_{n})\right)\Delta\phi
+ \kappa \pi \, C_I(\phi_{n})
\,, \label{Eab4}
\end{eqnarray}
where $\phi_{n+1} = \phi_{n} + \Delta\phi$.

Assuming that for a given $\phi_n$ the point $(C_I(\phi_n)=0,D_I(\phi_n))$
belongs to the orbit $ \Pi_{D_I}^I$,  the resulting point
$(C_I(\phi_{n+1}) = 0,D_I(\phi_{n+1}) = D_I(\phi_{n}))$ also belongs to the same
orbit $ \Pi_{D_I}^I$ independently of $\Delta\phi$.

These single-point orbits separate classical trajectories into $C_I>0$ and
$C_I<0$ regions. In fact, the region to which belong given trajectory depends on
sign of $C_I(\phi_0)$ of the initial conditions $(C_I(\phi_0)=0,D_I(\phi_0))$.
This is shown in the next section.

The space $S$ (see \eqref{cr1}) consists of the nonhyperbolic critical points so
that the neighbourhood of each such point includes rapidly changing
trajectories. Thus, the trajectories approaching asymptotically the orbits
$\Pi_{D_I}^I $ are very sensitive to the choice of the initial data for the
dynamics. This can be seen in Figs. \!\ref{fig:d1c1phic} and \ref{fig:d1c1phi}
(for the case $I = 1$).  These neighbourhoods represent the strange spikes.

It is expected that quantization may smear out the regions around the
single-point orbits so that the corresponding spikes may become more smooth.
 %color

%%%%%%%%%%%%%%%%%%%%%%%%%%%%%%%%%%%%%%%%%%%%%%%%%%%%%%%%%%%%%%%%%%%%%%%%%%%%%%%

\subsection{Classical spikes}

\subsubsection{Parametrization of dynamics by a scalar field}

In order to derive the spike solutions, within the dynamics parameterized by the
scalar field $\phi$, we follow the approach presented in Sec. II of
Ref. \cite{Czuchry:2016rlo}.

Let us assume that the initial conditions for $D_I$ and $C_I$ at $\phi = \phi_0$
have the form: $D_1 < D_2 < D_3 < 0$ and $1 \gg C_I > 0$.  Then, it follows from
\eqref{ab3}--\eqref{ab4} that $C_2$ and $C_3$ almost instantly vanish, while
$D_2$ and $D_3$ turn out to be essentially constant.  For later convenience we
define $D_\pm := D_2 + D_3 \pm 2 \sqrt{D_2 D_3}$. Therefore, the problem
reduces to finding the evolution of $C_1$ and $D_1$, which is governed by the
equations (here we denote by prime the derivative with respect to $\phi$):
\begin{align}
\kappa\pi C_1^\prime &= 4C_1 (-D_1 + D_2 + D_3)\,, \label{eq:2b.01}\\
\kappa\pi D_1^\prime &= C_1^2\,, \label{eq:2b.02}\\
-C_1^2 &= 4 (D_1 - D_+) (D_1 - D_-) - \kappa\pi^2\,, \label{eq:2b.03}
\end{align}
where the last one results from the constraint (\ref{b7}). Inserting the
right-hand side of (\ref{eq:2b.03}) into (\ref{eq:2b.02}), we obtain an equation
independent of $C_1$, whose solution can be written as
\begin{align}
D_1(\phi) &= D_2 + D_3 + \frac{1}{2} \sqrt{16 D_2 D_3 + \kappa \pi^2}\,
\tanh\left(\frac{2}{\kappa \pi} \sqrt{16 D_2 D_3 + \kappa \pi^2}\,
(\phi - \phi_0)\right. \nonumber\\
&\left.- {\rm arctanh}\sqrt{\frac{16 D_2 D_3 - C_{10}^2 + \kappa
\pi^2}{16 D_2 D_3 + \kappa \pi^2}}\right)\,. \label{eq:2b.04}
\end{align}
In the above expression the initial condition $D_1(\phi_0) = D_{10}$ has been
replaced by
\begin{align}
D_1(\phi_0) := D_{10} = D_2 + D_3 - \frac{1}{2} \sqrt{16 D_2 D_3 - C_{10}^2 + \kappa \pi^2}\,, \label{eq:2b.05}
\end{align}
due to the relation (\ref{eq:2b.03}) for $C_1(\phi_0) = C_{10}$. Furthermore, (\ref{eq:2b.04}) and (\ref{eq:2b.03}) give
\begin{align}
C_1(\phi) &= {\rm sgn}(C_{10}) \sqrt{16 D_2 D_3 + \kappa \pi^2}\, {\rm sech}\left(\frac{2}{\kappa \pi} \sqrt{16 D_2 D_3
+ \kappa \pi^2}\, (\phi - \phi_0)\right. \nonumber\\
&\left.- {\rm arctanh}\sqrt{\frac{16 D_2 D_3 - C_{10}^2 + \kappa \pi^2}{16 D_2 D_3 + \kappa \pi^2}}\right)\,, \label{eq:2b.06}
\end{align}
where ``${\rm sgn}$'' denotes the sign function (its value for $C_{10} = 0$ is irrelevant since then $C_1(\phi) = 0$). One can verify that
(\ref{eq:2b.06}) together with (\ref{eq:2b.04}) solve the equations
\eqref{eq:2b.01} and \eqref{eq:2b.03}.

\begin{figure}[h]
\includegraphics[width=0.45\textwidth]{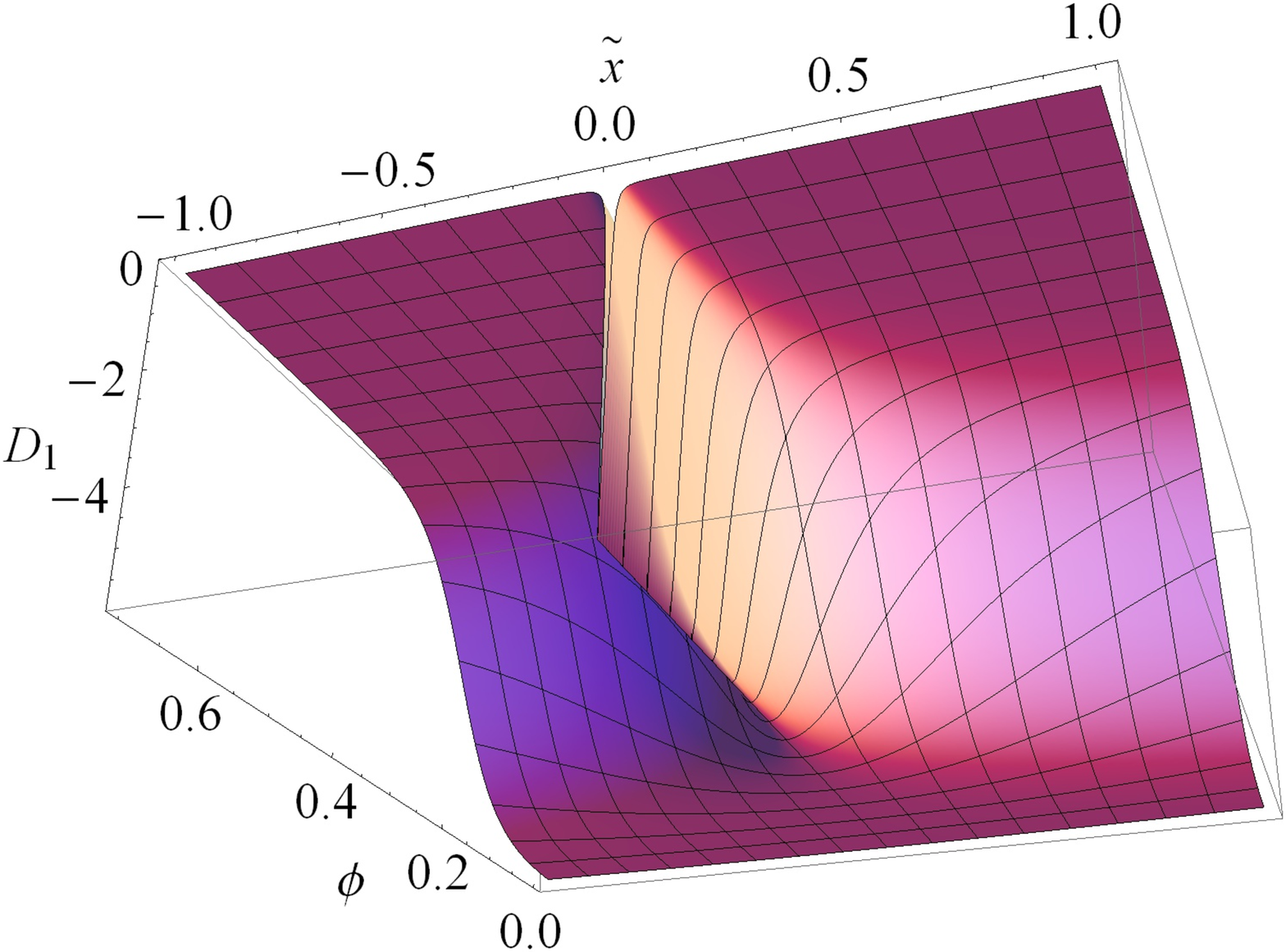}
\hspace{0.05\textwidth}
\includegraphics[width=0.45\textwidth]{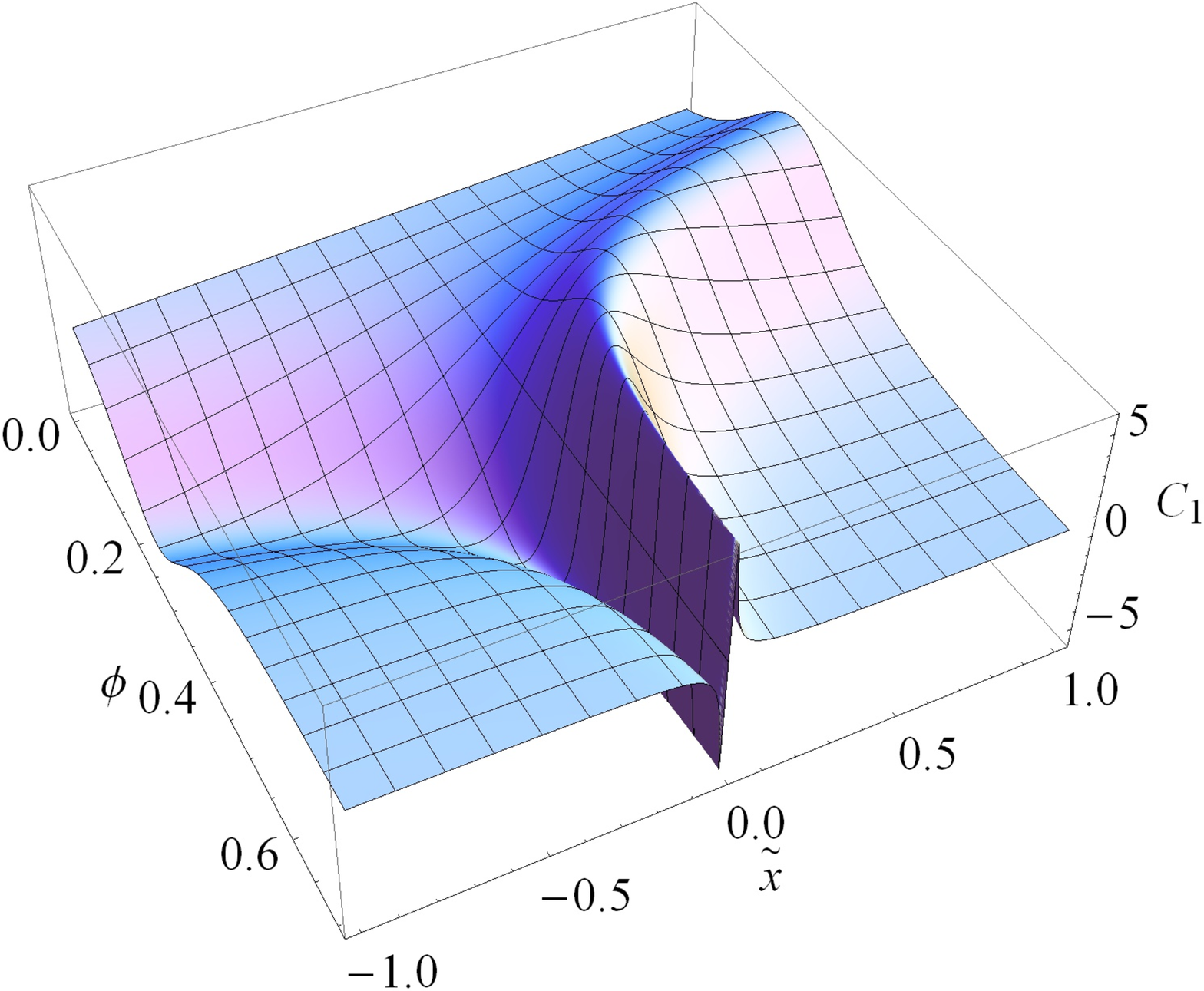}
\caption{\label{fig:d1c1phic} $D_1(\tilde x,\phi)$ (left) and $C_1(\tilde x,\phi)$ (right) as functions of two variables; lines of constant $\tilde x$ correspond to orbits (\ref{b8a})--(\ref{b8c}) of the affine group}
\end{figure}
\begin{figure}[h]
\includegraphics[width=0.45\textwidth]{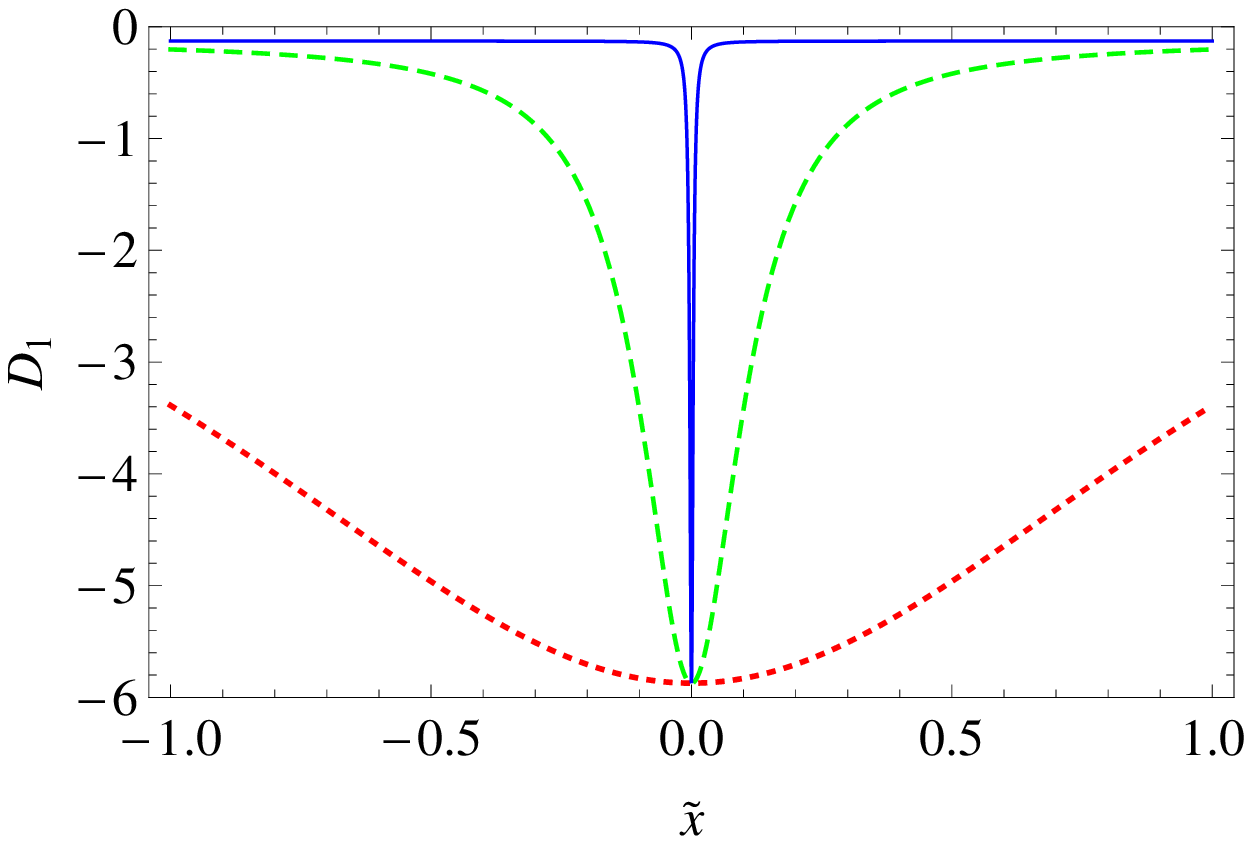}
\hspace{0.05\textwidth}
\includegraphics[width=0.45\textwidth]{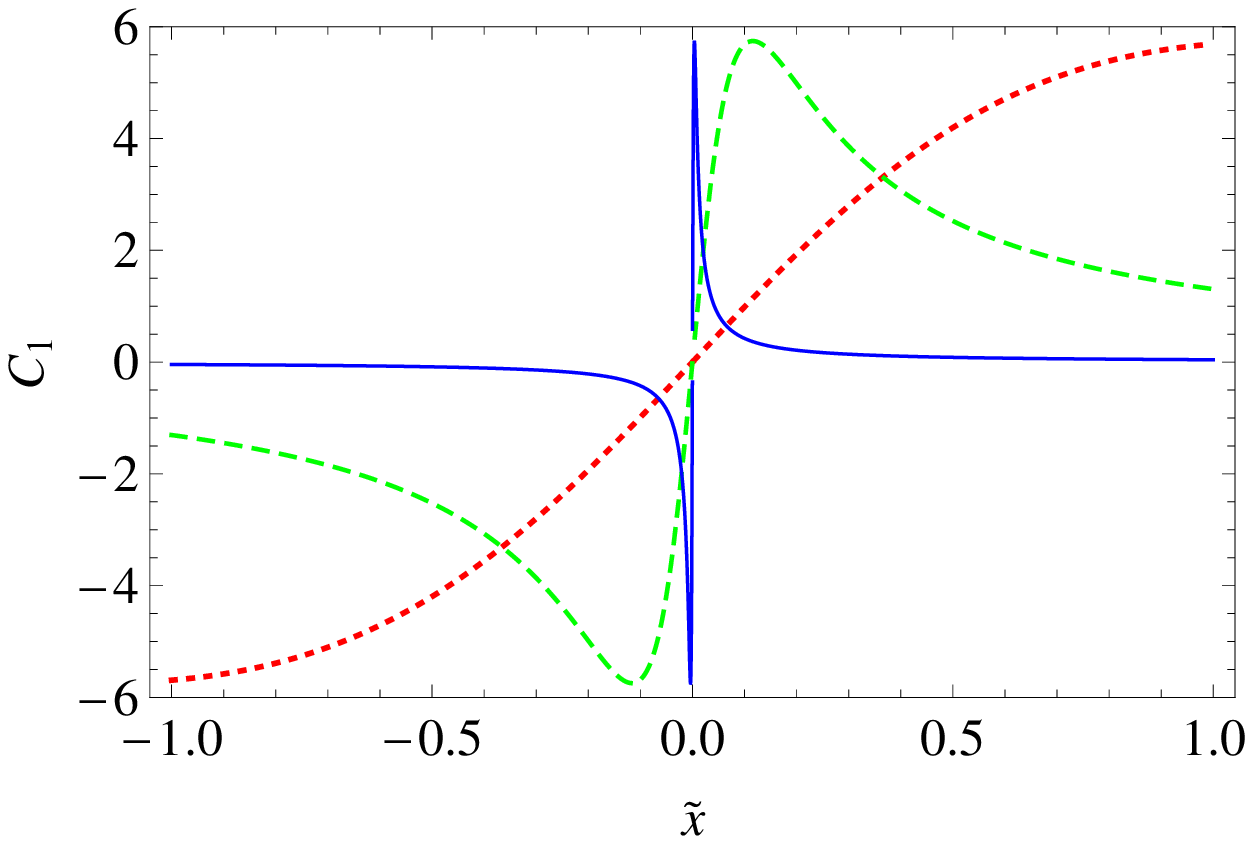}
\caption{\label{fig:d1c1phi} $D_1(\tilde x, \phi)$ (left) and $C_1(\tilde x, \phi)$ (right) as functions of $\tilde x$ for fixed values of $\phi = 0.2$ (red, dotted), $\phi = 0.4$ (green, dashed) and $\phi = 0.7$ (blue, solid); these are cross-sections of plots from Fig.~\ref{fig:d1c1phic}}
\end{figure}

Choosing the simple parametrization $C_{10} = \tilde x$, we can now draw $D_1$ and $C_1$ as functions
of both the evolution parameter $\phi$ and the initial condition $\tilde x$, or as functions of only $\tilde x$, for different fixed
values of $\phi$. Figs. \!\ref{fig:d1c1phic} and \ref{fig:d1c1phi} present the corresponding plots for
the setting of other quantities: $D_2 = -2$, $D_3 = -1$,
$\kappa = 1$, $\pi = 1$ and $\phi_0 = 0$. One can see that $D_1(\phi)$ and $C_1(\phi)$ behave in the
same way as $P_1(t)$ and $C_1(t)$
presented in Ref. \!\cite{Czuchry:2016rlo}, which is expected as the evolution parameter $\phi$ is a
monotonic function of the evolution
parameter $t$ owing to Eqs. \!\eqref{b5}--\eqref{b6}.

%%%%%%%%%%%%%%%%%%%%%%%%%%%%%%%%%%%%%%%%%%%%%%%%%%%%%%%%%%%%%%%%%%%%%%%%%%
\subsubsection{Parametrization of dynamics by the arc length}

\begin{figure}[h!]
\includegraphics[width=\textwidth]{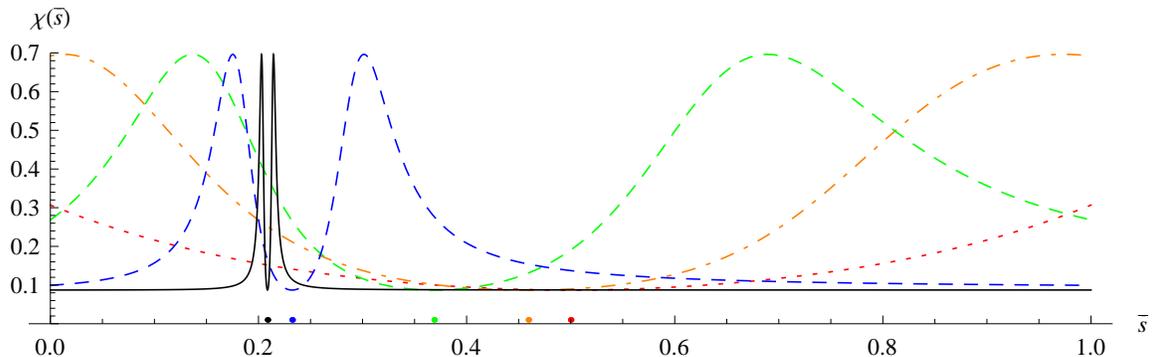}
\caption{\label{fig:chis} The generalized curvature $\chi(\bar s)$ of $(C_1(\bar
  s),D_1(\bar s))$ for evolution parameters $\phi = 0$ (red, dotted), $\phi =
  0.05$ (orange, dot-dashed), $\phi = 0.1$ (light green, dashed), $\phi = 0.2$
  (dark blue, dashed), and $\phi = 0.4$ (black, solid)}
\end{figure}

The arc length of the curve $\vec{r}(\tilde x) \equiv (C_1(\tilde x),D_1(\tilde x))$ is given by the
integral
\begin{align}
s(\tilde x) = \int_{\tilde x_0}^{\tilde x} dy\, \sqrt{\left(\frac{dC_1(y)}{dy}\right)^2 + \left(\frac{dD_1(y)}{dy}\right)^2}\,,
\label{eq:2b.07}
\end{align}
where $\tilde x_0$ is a certain chosen minimal value of $\tilde x$. Calculating
(\ref{eq:2b.07}) we obtain
\begin{align}
s(x) &= -\frac{1}{2} \sqrt{16 D_2 D_3 + \kappa \pi^2} \left(i\, {\rm E}(i\, \zeta,4) - i\, {\rm F}(i\, \zeta,4)
+ \sqrt{2 \cosh(2\zeta) - 1}\, \tanh\zeta\right)\,, \nonumber\\
\zeta &\equiv \frac{2}{\kappa \pi} \sqrt{16 D_2 D_3 + \kappa \pi^2}\, (\phi - \phi_0) - {\rm arctanh}\frac{\tilde x}{\sqrt{16 D_2 D_3
+ \kappa \pi^2}}\,,  \label{eq:2b.08}
\end{align}
where ${\rm F}$ denotes the elliptic integral of the first kind and ${\rm E}$ of
the second kind. This allows us to express the curve $\vec{r}(\tilde x)$ as a function of $s$, which needs to be calculated
numerically. Introducing the (normalized) Frenet vectors
\begin{align}
  \hat{e}_1(s) := \frac{1}{|\vec{e}_1(s)|}\, \vec{r}^{\, \prime}(s)\,, \qquad
  \hat{e}_2(s) := \frac{1}{|\vec{e}_2(s)|} \left(\vec{r}^{\, \prime\prime}(s) -
    \vec{r}^{\, \prime\prime}(s) \cdot \vec{e}_1(s)\,
    \vec{e}_1(s)\right)\,, %\label{eq:2b.09}
\end{align}
one can define the generalized curvature of $\vec{r}(s)$ as follows (see, e.g.,
\cite{WK})
\begin{align}
  \chi(s) := \frac{1}{|\vec{r}^{\, \prime}(s)|}\, \hat{e}_1^{\, \prime}(s) \cdot
  \hat{e}_2(s)\,. \label{eq:2b.10}
\end{align}
In Fig. \!\ref{fig:chis} we depict the generalized curvature of the curve
$(C_1(s),D_1(s))$ as a function of the normalized arclength $\bar s$
corresponding to $\tilde x \in [-5,5]$ (i.e. $s$ divided by the maximal value
$s(\tilde x = 5)$, for a given $\phi$) for different values of the evolution
parameter $\phi$. The values of $\kappa$, $\pi$, $\phi_0$ and $D_2$, $D_3$ are
kept the same as in the previous subsection. Moreover, dots on the horizontal
axis denote the value of $\bar s(\tilde x = 0)$ for a given $\phi$, which
naturally coincides with the middle of the spike.  The double peak corresponds
to the two inflection points of the curve visible on the right plot in
Fig. \!\ref{fig:d1c1phi}.

Fig. \!\ref{fig:chis} shows that the spike is created at some moment in the
evolution of the gravitational system and seems to be permanent. The shape of
the spike depends on time and changes from a plateau to singular structure.

%%%%%%%%%%%%%%%%%%%%%%%%%%%%%%%%%%%%%%%%%%%%%%%%%%%%%%%%%%%%%%%%%%%%%%%%%%%%%%
\section{Quantum level}

%%%%%%%%%%%%%%%%%%%%%%%%%%%%%%%%%%%%%%%%%%%%%%%%%%%%%%%%%%%%%%%%%%%%%%%%%%%%%%%
\subsection{Representation of the affine group}

The quantum version of the Lie algebra \eqref{b1} is defined by the algebraic
quantization principle: $C_I \rightarrow \hat{C}_I$ and $D_I \rightarrow
\hat{D}_I$, such that\footnote{Throughout the paper we choose $\hbar = 1$ and use
Planck's units except where otherwise stated.}
\begin{equation}\label{b88}
[\hat{C}_I, \hat{C}_J] = 0 = [\hat{D}_I, \hat{D}_J],
~~~~[\hat{C}_J, \hat{D}_I] = i \;\delta^I_J \hat{C}_I \; .
\end{equation}
where $I, J = 1, 2, 3$.
The commutation relations  \eqref{b88} are the same as for the generators of the
affine group \cite{AslaksenKlauder:1968}.

The affine group $\Group{Aff}(\dR_+)_I$ generated by the pair $\hat{C}_I$ and
$\hat{D}_I$ has two inequivalent unitary representations $U_{-}(p,q)_I$ and
$U_{+}(p,q)_I$. They are constructed in two carrier spaces of square integrable
functions $L^2(\dR_-,d\nu(x^I))$ and $L^2(\dR_+,d\nu(x^I))$,
$d\nu(x^I)=dx^I/|x^I|$, which correspond to the negative and positive
spectrum of the position operator $\hat{C}_I$, respectively. Because of physical
interpretation we needs the full spectrum of the position operator. This
requirement enforces using the reducible representation of the affine group in
the carrier space $\mathcal{K}_I:=L^2(\dR_-,d\nu(x^I)) \oplus
L^2(\dR_+,d\nu(x^I))$. The general form of the vector $f \in \mathcal{K}_I$ can
be written as a direct sum of the functions $f_{\mp} \in L^2(\dR_\mp,d\nu(x^I)$:
\begin{equation}
\label{VectPlusMinus}
f=f_{-} \oplus f_{+} \, .
\end{equation}
The scalar product of such two vectors is the sum of the appropriate partial scalar products:
\begin{align}
\BraKet{f_1 \oplus f_2}{g_1 \oplus g_2} &:=
\BraKet{f_1}{g_1}_- + \BraKet{f_2}{g_2}_+ \nonumber\\
&= \int_{-\infty}^0 d\nu(x^I) f_1(x^I)^\star g_1(x^I) +
\int_0^\infty d\nu(x^I) f_2(x^I)^\star g_2(x^I)\,. \label{ScalProdSimpleSum}
\end{align}
The action of the  affine group $\Group{Aff}(\dR_+)_I$ in this carrier space
$\mathcal{K}_I$ can be written as
\begin{equation}
\label{AffineActionKI}
U(p,q)_I f=U_{-}(p,q)_I f_{-} \oplus U_{+}(p,q)_I f_{+},
\end{equation}
where $p \in \dR$, $q \in \dR_+$ and
\begin{equation}
\label{AffineActionKI2}
U_{\mp}(p,q)_I f_{\mp}(x^I) = e^{ipx^I} f_{\mp}(qx^I).
\end{equation}
This structure allows for extension of this affine action to the whole straight
line.  For this purpose it is enough to extend the
appropriate functions from half-line to the full straight line: $f_{-}(x^I)=0$
for $x^I \geq 0$ and $f_{+}(x^I)=0$ for $x^I \leq 0$. Then, denoting by $\Ket{x^I
  \oplus x^I}$ the ``position'' vector in the space $\mathcal{K}_I$, every
function belonging to $\mathcal{K}_I$ can be represented as:
\begin{equation}
\label{PositionRepFun}
f(x^I):=\BraKet{x^I \oplus x^I}{f_{-} \oplus f_{+}}=
\BraKet{x^I}{f_{-}} + \BraKet{x^I}{f_{+}}= f_{-}(x^I) + f_{+}(x^I).
\end{equation}
It is obvious that the space $\mathcal{K}_I \subset L^2 (\dR,d\nu{(x^I)})$ and that
the scalar product \eqref{ScalProdSimpleSum} can be rewritten as
\begin{equation}
\label{ScalProdSimpleSum2}
\BraKet{f_1 \oplus f_2}{g_1 \oplus g_2}_I = \BraKet{f}{g}_I=
\int_{-\infty}^{\infty} d\nu(x^I) f(x^I)^\star g(x^I)\,.
\end{equation}
The action of the  affine group $\Group{Aff}(\dR_+)_I$ in this new carrier space,
which we denote again by $\mathcal{K}_I$, can be written as
\begin{equation}
\label{AffineActionKI2}
U(p,q)_I f(x^I)= e^{ipx^I} f(qx^I)\,.
\end{equation}
The explicit representation of the generators of this group are given by the following
operators
\begin{equation}\label{c1}
\hat{D}_I f(x^I) := -i\, x^I \frac{\partial}{\partial x^I}\, f(x^I)\,,
\qquad \hat{C}_I f(x^I) := x^I f(x^I)\,,
\end{equation}
where $I = 1,2,3$.

The corresponding unitary operators representing elements of the affine group
are:
\begin{equation}\label{UnOperAffGrp}
\hat{U}(p,q)_I= e^{ip\hat{C_I}} e^{i\ln(q)\hat{D_I}}
\end{equation}
where $-\infty < p < +\infty$, $0 < q < +\infty$.

Taking into account three variables $x^I$ $(I=1,2,3)$,
the carrier space  $\mathcal{K}$ for the representation of the algebra
\eqref{b88} can be defined to be
\begin{equation}\label{car1}
\mathcal{K}:=
\mathcal{K}_1\otimes \mathcal{K}_2\otimes  \mathcal{K}_3 \, ,
\end{equation}
where
\begin{equation}\label{car2}
\mathcal{K}_I = L^2(\dR_{-}, d\nu(x^I)) \oplus   L^2(\dR_{+}, d\nu(x^I))
\subset  L^2 (\dR, d\nu(x^I))
\end{equation}%
and the scalar product is constructed according to prescription for tensor
product of Hilbert spaces:
\begin{equation}
\label{ScalarProductTotal}
\BraKet{f}{g} =
\int_{-\infty}^{\infty} d\nu(x^1)
\int_{-\infty}^{\infty} d\nu(x^2)
\int_{-\infty}^{\infty} d\nu(x^3)
f(x^1,x^2,x^3)^\star g(x^1,x^2,x^3)\,.
\end{equation}
The ``total'' affine group used in this paper is the direct product of the three affine
groups $\Group{Aff}_0=\Group{Aff}(\dR_{+})_1 \otimes \Group{Aff}(\dR_{+})_2
\otimes \Group{Aff}(\dR_{+})_3$.
This realization of the affine group allows for physical interpretation of
quantized $C_I$ and $D_I$ variables.

%%%%%%%%%%%%%%%%%%%%%%%%%%%%%%%%%%%%%%%%%%%%%%%%%%%%%%%%%%%%%%%%%%%%%%%%%%
\subsection{Quantum dynamics}

The quantum dynamics of our system may be derived, to some extent, from the
quantum version of the Hamiltonian constraint defined by Eq. \!\eqref{b7}. In a
standard approach, one maps the dynamical constraint into an operator defined in
kinematical Hilbert space. Its kernel may be used to construct physical Hilbert
space. However, such an approach leads to the problem of time at the quantum
level.

The reason for having the scalar field in the Hamiltonian \eqref{b7}, is the
hope that it may resolve the problem of time both at classical and quantum
levels. Such an approach works in the classical case as it leads to the relative
dynamics, defined by Eqs. \!\eqref{ab3}--\eqref{ab4}, parameterized by the
scalar field $\phi$. However, an extension of this strategy to the quantum level
faces serious difficulty.  Namely, quantization of the scalar field algebra
$\{\phi,\pi\}= 1$ as follows
\begin{equation}\label{c2}
\hat{\pi} f (\phi) :=
-i \frac{\partial}{\partial \phi} f(\phi),
~~~ \hat{\phi}f(\phi):=  \phi f(\phi),~~~ f \in L^2 (\dR, d\phi) \, ,
\end{equation}
so that $[\hat{\phi},\hat{\pi}] = i \id $, leads to the inconsistency.
In this case, according to standard approach to quantum mechanics,
  $\phi$ represents an additional degree of freedom of our quantum system.  The
  field $\phi$ is the variable involved in every required wave function and it
  cannot be considered as a parameter representing a reference clock we want to
  introduce. One needs to notice that every quantum amplitudes is independent of
  the variable $\phi$ because they are obtained by calculating the appropriate
  scalar product containing among others integration over $\phi$.

  To parameterize with $\phi$ the reference classical clock uncoupled to our
  quantum system one needs to construct a hybrid approximation of deterministic
  unitary quantum evolution: the field should evolves in a classical way and
  quantum states of the system should evolve according to a unitary
  prescription.

  Let us treat the field $\phi$ as a classical field which value is considered
  as a parameter showing tics of a classical clock, that is $\phi$ is a
  parameter enumerating changes of our Hamiltonian system.

  We propose to modify Schr\"odinger type unitary evolution operator to the form
  containing both: evolution of the classical field and evolution of the quantum
  system itself. This operator we denote by $\mathcal{U}(\phi,\phi_0)$. It is
  defined by a series of natural conditions:
\begin{itemize}
\item First of all, the operator $\mathcal{U}(\phi,\phi_0)$ evolves the quantum
  state of our gravitational system from the ``time'' $\phi_0$ and the state
  $\Psi_1$ to the ``time'' $\phi$ and the state $\Psi_2$ as follows
\begin{equation}
\label{EvolOperDef}
\mathcal{U}(\phi,\phi_0) \Psi_1(\phi_0,x_1,x_2,x_3)
= \Psi_2(\phi,x_1,x_2,x_3)\,,
\end{equation}
where $\Psi_1\, \Psi_2 \in \mathcal{K}$, and where $\mathcal{K}$ is a Hilbert
space.  Thus $\mathcal{U}$ changes the state vector in the Hilbert state space
and the time parameter by changing the field.

Since the time parameter $\phi$ does not couple to the gravitational field in
\eqref{b7}, we can factorize the evolution operator $\mathcal{U}(\phi,\phi_0)$
into two independent operations:  \\
a) the unitary operator $V_\mathcal{K}(\phi,\phi_0)$ acting on the
spatial dependance of state vectors in the Hilbert space $\mathcal{K}$ while the
field $\phi$ is changing, \\
b) the operation $V_{\pi}(\phi,\phi_0)$ acting on the parametric
dependence of the state vectors of the field $\phi$.

In what follows, we assume the dependence of the evolution operator on the
difference $\tau = \phi - \phi_0$ between the final and initial value of the
field $\phi$, i.e. we assume the translational invariance of the evolution
operator with respect to the parameter $\phi$.  This means that $\mathcal{U}$
does not depend on the choice of the initial time $\phi_0$, but only on $\tau$.
Thus, the full evolution operator can be written as
\begin{equation}
\label{EvolOpFactorization}
\mathcal{U}(\tau) = V_\mathcal{K}(\tau) V_\pi(\tau)\,.
\end{equation}

\item The evolution operator fulfils the standard conditions for quantum
  evolution:
\begin{eqnarray}
\label{EvolOperatorsConditions}
&& \mathcal{U}(0)=\UnitOp \quad \mbox{ (no shift in ``time'') }, \\
&& \mathcal{U}(\tau_2+\tau_1) =\mathcal{U}(\tau_2)\mathcal{U}(\tau_1)
\quad \mbox{ (no ``holes'' in the evolution) }, \\
&& \mathcal{U}(\tau)^{\dagger}=\mathcal{U}(\tau)^{-1}=\mathcal{U}(-\tau)
\quad \mbox{ (unitarity) }.
\end{eqnarray}
The first one represents the fact, that if there is no shift in time, the state vector
stays the same. The second means that every evolution can be split
into intermediate steps. These two conditions are expected to hold for both the
classical end quantum evolution.  The last line represents the unitarity
condition which is related to probabilistic interpretation of quantum mechanics.

To fulfil the last condition the parametric part of the evolution operator has
to transform as the complex conjugation:
\begin{equation}
\label{EvolOperatorsConditions2}
[V_{\mathcal{K}}(\tau) V_{\pi}(\tau)]^\dagger =
V_{\mathcal{K}}(\tau)^\dagger V_{\pi}(\tau)^\star
\end{equation}
\end{itemize}
 %color
%
Let us now consider a formal shift operation with respect to the field $\phi$.
For this purpose we define a kind of adjoint action of the field $\phi$ and its
canonically conjugate momentum $\pi$ on the classical phase space. For an
arbitrary function $g(\phi,\pi)$ on this phase space the adjoint action is
defined to be
\begin{align}
\label{ShiftOperPhaseSpDef}
\{h(\phi,\pi),\cdot\} f(\phi,\pi) &:= \{h(\phi,\pi),f(\phi,\pi)\}\,,
\nonumber \\
\{\cdot,h(\phi,\pi)\} f(\phi,\pi) &:= \{f(\phi,\pi),h(\phi,\pi)\}\,,
\nonumber \\
\{h(\phi,\pi),\cdot\} &= -\{\cdot,h(\phi,\pi)\}\,,
\end{align}
where the Poisson bracket is given by
\begin{equation}
\label{PoissonBracketDef}
\{h(\phi,\pi),f(\phi,\pi)\} :=
\frac{\partial h(\phi,\pi)}{\partial \phi}
\frac{\partial f(\phi,\pi)}{\partial \pi}
- \frac{\partial h(\phi,\pi)}{\partial \pi}
\frac{\partial f(\phi,\pi)}{\partial \phi}\,.
\end{equation}
One can directly check that
\begin{equation}
\label{ShiftPhaseSpOp}
e^{\tau \{\cdot,\pi\}} f(\phi,\pi) =
e^{-\tau \{\pi,\cdot\}} f(\phi,\pi) = f(\phi+\tau,\pi)\,,
\end{equation}
where
\begin{equation}
\label{ShiftPhaseSpOpSeries}
e^{\tau \{\pi,\cdot\}} =
\sum_{n=0}^\infty \frac{\tau^n \{\pi,\cdot\}^{(n)}}{n!}\,.
\end{equation}
The powers of the adjoint action are understood as
\begin{equation}
\label{PwersAdjAct}
\{\pi,\cdot \}^{(n)} f(\phi,\pi) = \underbrace{
\{\pi,\{\pi,\dots\{\pi,\{}_n\pi,f(\phi,\pi)\}\dots\}\,,
\end{equation}
where $\{\pi,\cdot\}^{(0)}f(\phi,\pi) = f(\phi,\pi)$.

The isomorphic realization of the classical shift operation
  \eqref{ShiftPhaseSpOp} with respect to the field $\phi$ in the state space is
  given by
\begin{equation}
\label{eq:CLShiftQM}
e^{\tau \{\pi,\cdot\}}  \to e^{\tau \frac{\partial}{\partial\phi}} \  .
\end{equation}
As a consequence the formal shift of the state $\Psi(\phi,x)$ in respect to the
time $\phi$ is given by
\begin{equation}
\label{ShiftOperHilbertSpDef}
\Psi(\phi + \tau,x) = e^{\tau \frac{\partial}{\partial\phi}} \Psi(\phi,x)\,,
\end{equation}
where $x := (x_1,x_2,x_3)$.
%
%color

The comparison of the operations \eqref{ShiftOperHilbertSpDef} and
\eqref{ShiftPhaseSpOp} suggests that the shift generator
$\frac{\partial}{\partial \phi} =: \check{\pi}$, defined in the quantum state
space, may play a similar role to the classical momentum $\pi$ acting (by the
adjoint action) in the phase space. Working in the quantum state space we
postulate the replacement of the classical momentum $\pi$ with the operation
$\check{\pi}$.

The classical evolution in the phase space can be written in terms of
  the adjoint action as $e^{\tau \{E(\pi),\cdot\}}$, where $E(\pi)$ is a real
  function of the momentum $\pi$ generating evolution of this free field.  As a
  consequence of \eqref{eq:CLShiftQM} the appropriate realization of this
operation to be a part of the evolution operator is the replacement $E(\pi) \to
-iE(\check{\pi})$. The imaginary unit has to be added to fulfil the unitarity
requirement \eqref{EvolOperatorsConditions2}. The classical part of the
evolution operator is expected to be:
\begin{equation}
\label{EvolOpParameterPhi}
V_\pi(\tau) = e^{-i \tau E(\check{\pi})},
\end{equation}
%
%
%color
where $E(\check{\pi})$ is some real function of $\check{\pi}$.

Making use of \eqref{ShiftOperHilbertSpDef}--\eqref{EvolOpParameterPhi} and the
factorization \eqref{EvolOpFactorization}, we rewrite \eqref{EvolOperDef} as
follows
\begin{equation}
\label{StateEvolutionU}
e^{\tau \check{\pi}} \Psi(\phi,x)
= V_\mathcal{K}(\tau) e^{-i \tau E(\check{\pi})} \Psi(\phi,x)
\end{equation}
Taking derivative of \eqref{StateEvolutionU} with respect to $\tau$, at
$\tau = 0$, leads to the local evolution equation:
\begin{equation}
\label{StateEvolutionEq}
\check{\pi} \Psi(\phi,x)=
\left[ \left( \frac{\partial V_{\mathcal{K}}(\tau)}{\partial \tau}
\right)_{\tau=0}
- i E\left(\check{\pi} \right) \right] \Psi(\phi,x)\, .
\end{equation}
%
 % Red
Introducing
\begin{equation}
\label{Mhamiltonian}
\hat{W}:=i \left( \frac{\partial V_{\mathcal{K}}(\tau)}{\partial \tau}
\right)_{\tau=0} \, ,
\end{equation}
we can rewrite  \eqref{StateEvolutionEq} in the form
\begin{equation}
\label{StateEvolutionEq2}
i\frac{\partial \Psi(\phi,x)}{\partial\phi}
= \left[ \hat{W} +
E \left(\frac{\partial}{\partial \phi}\right) \right] \Psi(\phi,x).
\end{equation}
Assuming
\begin{equation}
\label{SeparationVarEvolEq}
\Psi(\phi,x) = \omega(\phi) \psi(x)\, ,
\end{equation}
enables rewriting \eqref{StateEvolutionEq2} in the separable form
\begin{equation}
\label{StateEvolutionEq3}
\frac{1}{\omega(\phi)}\left[ i\frac{\partial}{\partial\phi}
-E\left(\frac{\partial}{\partial \phi}\right) \right] \omega(\phi)
= \frac{1}{\psi(x)} \hat{W} \psi(x) \,,
\end{equation}
which leads to the two eigenequations:
\begin{equation}
\label{StEvolEigenEq}
\left[ i\frac{\partial}{\partial\phi}
-E\left(\frac{\partial}{\partial \phi}\right) \right] \omega_{\lambda}(\phi)
= \lambda \omega_{\lambda}(\phi) \,,
\end{equation}
and
\begin{equation}
\label{StEvolEigenEq2}
\hat{W} \psi_{\lambda}(x) = \lambda \psi_{\lambda}(x).
\end{equation}

We assume, according to \eqref{EvolOpFactorization}, that the quantum evolution
operator corresponding to the classical constraint consists of the quantized
position dependent part of \eqref{b7} and the shifted parametric part of
\eqref{b7}. This way we avoid quantization of the algebra $\{\phi,\pi\}= 1$, and
consequently quantization of the classical time variable $\phi$. Both classical
and quantum evolutions are now parameterized by a single variable $\phi$ that we
call the time.

Since there are no products of $C_I$ and $D_I$ in \eqref{b7}, and due to
\eqref{b88}, the mapping of $H$ defined by \eqref{b7} into a Hamiltonian
operator $ \hat{H}$ is straightforward.  We get
\begin{eqnarray}\label{r2}
&& \hat{H} = E\left(\frac{\partial}{\partial \phi}\right) + \hat{W}
\\
&&= 2 \left(-\frac{\kappa}{4} \frac{\partial^2}{\partial\phi^2} +
\sum_I x_I^2 \frac{\partial^2}{\partial x_I^2} -
2 \sum_{I<J} x_I x_J \frac{\partial^2}{\partial x_I \partial x_J} +
\sum_I x_I \frac{\partial}{\partial x_I} \right)
+ \sum_{I<J} x_I x_J - \frac{1}{2} \sum_I x_I^2\,, \nonumber
\end{eqnarray}
which implies that
\begin{eqnarray}
&&\hat{W} = 2 \left(\sum_I x_I^2\frac{\partial^2}{\partial x_I^2} -
2 \sum_{I<J} x_I x_J \frac{\partial^2}{\partial x_I \partial x_J} +
\sum_I x_I \frac{\partial}{\partial x_I} \right)
+ \sum_{I<J} x_I x_J - \frac{1}{2} \sum_I x_I^2\,, \label{WEO1}\\
&&E \left(\frac{\partial}{\partial\phi}\right)
= -\frac{\kappa}{2} \frac{\partial^2}{\partial\phi^2}\,. \label{WEO2}
\end{eqnarray}
%

%%%%%%%%%%%%%%%%%%%%%%%%%%%%%%%%%%%%%%%%%%%%%%%%%%%%%%%%%%%%%%%
\subsubsection{Solving the eigenequation  \eqref{StEvolEigenEq} analytically}
%%%%%%%%%%%%%%%%%%%%%%%%%%%%%%%%%%%%%%%%%%%%%%%%%%%%%%%%%%%%%%%

Making use of  \eqref{WEO2}, we get \eqref{StEvolEigenEq} in the form
\begin{equation}
\label{SecEigenEqPhi}
\left( i\frac{d}{d\phi} + \frac{\kappa}{2}\frac{d^2}{d\phi^2} \right)
\omega_{\lambda}(\phi) = \lambda \omega_{\lambda}(\phi) \, .
\end{equation}
The solution to \eqref{SecEigenEqPhi}, for $\kappa\lambda \neq 1/2$, is found to
be
\begin{equation}\label{solE1}
\omega_{\lambda}(\phi) = e^{-i\kappa\phi} \{A_\lambda \exp(\kappa\sqrt{2\kappa\lambda -1}\,\phi)
+ B_\lambda \exp(-\kappa\sqrt{2\kappa\lambda -1}\,\phi)\}\, ,
\end{equation}
whereas for $\kappa\lambda = 1/2$ one has
\begin{equation}\label{solE2}
\omega_{\lambda}(\phi) = (A_\lambda \phi + B_\lambda ) e^{-i\kappa\phi} \, ,
\end{equation}
where $A_\lambda$ and $B_\lambda$ are  arbitrary  constants. In what follows we denote the
solutions \eqref{solE1}--\eqref{solE2} as
$\omega_{\lambda}(A_\lambda,B_\lambda;\phi)$.

%%%%%%%%%%%%%%%%%%%%%%%%%%%%%%%%%%%%%%%%%
\subsubsection{Solving the eigenequation \eqref{StEvolEigenEq2} by variational method}\label{solvar}
%%%%%%%%%%%%%%%%%%%%%%%%%%%%%%%%%%%%%%%%%

The eigenequation \eqref{StEvolEigenEq2} can be solved numerically in terms of
given finite basis of functions $\{ \psi_n \}_{n=0}^{N}$,
by taking the solution $\psi_\lambda$ in the form
\be
\label{Nfun}
\psi_\lambda \simeq  \psi_\lambda^N = \sum_{n=0}^N c_n \psi_n \, ,
\ee
where $c_n$ are unknown coefficients to be determined. The functions $\psi_n$ should be consistent
with the boundary conditions.
It means, they should vanish sufficiently fast at zero and infinity to satisfy the condition
\begin{equation}\label{error2}
\parallel \psi_\lambda \parallel^2 = \int_{-\infty}^{\infty} \frac{d x_1}{|x_1|} \int_{-\infty}^{\infty}
\frac{d x_2}{|x_2|} \int_{-\infty}^{\infty}
\frac{d x_3}{|x_3|} | \psi_\lambda |^2 < \infty \, .
\end{equation}
The coefficients $c_n$ can be found by considering the following functional:
\be
\label{error}
R[ \psi_\lambda ] := \frac{\parallel \hat{W} \psi_\lambda - \lambda \psi_\lambda
\parallel^2}{\parallel \psi_\lambda  \parallel^2} \, .
\ee
It is clear that \eqref{error} vanishes identically if $\psi_\lambda^N $ is an exact solution
to the equation \eqref{StEvolEigenEq2}. If this is not the case but  $R[ \psi_\lambda ] \ll 1$,
then we have an approximate solution. The smaller $R[ \psi_\lambda ]$,
the better the approximation. The latter fact suggests a method of finding the numerical solution.
Namely, one can minimize \eqref{error}  with respect to all unknown coefficients, including the
eigenvalue $\lambda$. This fixes all the parameters
in Eq.  \!\eqref{Nfun} and determines the error $R[ \psi_\lambda ]$.

To start the procedure one should fix the basis $\{ \psi_n\}$. It is reasonable
to incorporate the fact that the operator $\hat{W}$ is invariant under $S_3$
group of permutations of the variables $\{ x_1,x_2,x_3\}$. Therefore, looking
for the basis it is reasonable to consider functions sharing this symmetry,
i.e. requiring they are symmetric with respect to the replacements $x_i
\leftrightarrow x_j$.  A convenient choice is provided by the following ansatz
\begin{align}
\nonumber
\label{ans1}
(\psi_S)^N_{\alpha} = |x_1 x_2 x_3|^\alpha \sum_{n_1+n_2+n_3 \le N} \frac{c_{(n_1 n_2 n_3)}}{(n_1+n_2+n_3)!}
\left( \ln x_1^2 \right)^{n_1} \left( \ln x_2^2 \right)^{n_2} \left( \ln x_3^2 \right)^{n_3} \cdot
\\[1ex]
\cdot \exp\left( -\frac{1}{2} (\gamma+ i \tilde{\gamma}) (|x_1|+|x_2|+|x_3|) \right),
\end{align}
where $\alpha \geq \frac{1}{2}$, $\gamma>0$, $\tilde{\gamma} \in \mathbb{R}$, while $\sum_{n_1+n_2+n_3 \le N}$
stands for the sum over $n_1,n_2,n_3 \in [0,N]$ such that $n_1+n_2+n_3 \le N$, i.e. the series \eqref{ans1} is
terminated at the $N$-th order ($N = n_1+n_2+n_3$). The bracket $(n_1, n_2, n_3)$ denotes ordering operation,
e.g.  $c_{(0 2 3)} = c_{0 2 3}$, $c_{(2 0 3)} = c_{0 2 3}$, etc. The operation guaranties that the function
$\psi^N_\lambda$ consists of symmetric terms with respect to the replacement $x_i \leftrightarrow x_j$.
For instance, there are two second-order $(N=2)$ terms in \eqref{ans1}: $\frac{1}{2}c_{011} ( \ln x_1^2
\ln x_2^2  + \ln x_1^2 \ln x_3^2 + \ln x_2^2 \ln x_3^2 )$ and $ \frac{1}{2}c_{002}
 (\left( \ln x_1^2 \right)^2+\left( \ln x_2^2 \right)^2+\left( \ln x_3^2 \right)^2)$. The additional weights
$1/(n_1+n_2+n_3)!$ are introduced for technical simplicity. They guarantee that the coefficients
$c_{n_1  n_2  n_3}$, fixed by the minimization procedure, are of a similar order. The latter improve the
minimization. Note that the number of $c_{n_1 n_2 n_3}$ grows fast with order $N$.

Taking $\alpha = \frac{1}{2}$ and $N=10$, one finds a solution $\psi_{\lambda_1} = (\psi_S)^{10}_{1/2}$
specified by numerical parameters $( \lambda,\gamma,\tilde{\gamma},R[ \psi_{\lambda_1}] ) = ( \lambda_1,
\gamma_1,\tilde{\gamma}_1,R_1 )$, where
\be
\label{nsol1}
\lambda_{1} \simeq -0.0821, \quad \gamma_1 \simeq 1.229, \quad \tilde{\gamma}_1 = -4.05 \times 10^{-4}
\quad R_1 \simeq 0.0172,
\ee
while the coefficients $c_{n_1 n_2 n_3}$ are given explicitly in Appendix \ref{num}.
The choice $\alpha = \frac{1}{2}$ leads to
the smallest global error $R_1$. The point-like precision defined as
\be
\label{localerror}
E[\psi_\lambda] := \sup_{x_I \in \mathbb{R}^3}
( |\hat{W} \psi_\lambda - \lambda \psi_\lambda|)
\ee
gives  $E[\psi_{\lambda_1}] \simeq 0.0028$. In  \eqref{localerror} $\psi_\lambda$ stands for
a normalized function.

For the second numerical solution we take the ansatz:
\be\label{ans2}
(\psi_A)^N_\alpha = \left(\textmd{sign}(x_1)+\textmd{sign}(x_2)+\textmd{sign}(x_3)\right) (\psi_S)^{N}_{\alpha}.
\ee
As the solution is antisymmetric,  it is orthogonal to the previous one, i.e. $\langle (\psi_A)^{N_1}_{\alpha_1}|
(\psi_S)^{N_2}_{\alpha_2} \rangle = 0$. Taking as before $\alpha = \frac{1}{2}$ and $N=10$,  we get the function
$\psi_{\lambda_2} = (\psi_A)^{10}_{1/2}$. Applying our method of fixing the coefficients in the ansatz leads to
\be
\label{nsol2}
\lambda_2 \simeq - 0.0957, \quad \gamma_2 \simeq 1.369, \quad \tilde{\gamma}_2 \simeq 4.46 \times 10^{-3},
\quad R_2 \simeq 0.0218.
\ee
The coefficients $c_{n_1 n_2 n_3}$ are listed in Appendix \ref{num}. As in the case of $\psi_{\lambda_1}$
the point-like precision
 \eqref{localerror} gives $ E[\psi_{\lambda_2}] \simeq 0.0058$.

%%%%%%%%%%%%%%%%%%%%%%%%%%%%%%%%%%%%%%%%%%
\subsubsection{Solving the eigenequation  \eqref{StEvolEigenEq2} by spectral method}\label{solspc}
%%%%%%%%%%%%%%%%%%%%%%%%%%%%%%%%%%%%%%%%%
We start with the ansatz
\be
\label{ansa1}
\psi = |x_1 x_2 x_3|^\alpha f(x_1,x_2,x_3) \exp \left(-\frac{\gamma}{2}(|x_1|+|x_2|+|x_3|) \right),
\ee
where $\gamma>0$, $\alpha \geq 1/2$. The eigenequation  \eqref{StEvolEigenEq2} can be rewritten as
\be
\hat{W}\psi - \lambda \psi =  |x_1 x_2 x_3|^\alpha \left( \hat{F}_{\alpha \gamma}(f(x_1,x_2,x_3))-\lambda
f(x_1,x_2,x_3) \right) \exp \left(-\frac{\gamma}{2}(|x_1|+|x_2|+|x_3|) \right),
\ee
where
\begin{align}
\hat{F}_{\alpha \gamma} &= 2 \sum_I x_I^2 \frac{\partial^2}{\partial x_I^2} - 4
\sum_{I<J} x_I x_J \frac{\partial^2}{\partial x_I \partial x_J} \nonumber\\
&+ 2 \sum_I \left(1 - 2\alpha + \gamma (|x_1| + |x_2| + |x_3| - 2 |x_I|)\right) x_I
\frac{\partial}{\partial x_I} \nonumber\\
&+ \sum_I \left(\frac{\gamma^2 - 1}{2}\, x_I^2 + \gamma (2\alpha - 1) |x_I|\right) +
\sum_{I<J} \left(x_I x_J +\gamma^2 |x_I| |x_J|\right) - 6\alpha^2\,. \label{Fform}
\end{align}

With the ansatz \eqref{ansa1}, the problem of solving the eigenequation \eqref{StEvolEigenEq2}
reduces to the problem of solving the corresponding eigenequation
for $\hat{F}$ operator, i.e.
\be
\label{Feq}
\hat{F}_{\alpha \gamma} f(x_1,x_2,x_3) = \lambda f(x_1,x_2,x_3)\,.
\ee
We solve  Eq. \eqref{Feq} by using the spectral methods \cite{Grandclement:2007sb}. This form
is much more convenient
from numerical point of view because of the lack of two terms $|x_1 x_2 x_3|^\alpha$ and
$\exp(-\gamma/2 (|x_1| + |x_2| + |x_3|))$,
causing additional numerical errors\footnote{More precisely, within the spectral method,
these terms result in combination of small and
large numbers (components of the matrix representing an approximate form of the eigenequation
at a lattice).}.

It is convenient, for numerical treatment, to assume

\be
\label{eff}
f(x_1,x_2,x_3) = \sum_{n_1=1}^N \sum_{n_2=1}^N \sum_{n_3=1}^N c_{n_1,n_2,n_3}
f_{n_1 n_2 n_3}(x_1,x_2,x_3)\,,
\ee
where $N>1$ is the cut-off, while $f_{n_1 n_2 n_3}$ stand for a fixed basis of functions.
The standard procedure involves cosine function, however, it will be convenient to adopt
a different choice.
The solution to the eigenequation \eqref{Feq} is specified by fixing the unknown coefficients
$c_{n_1,n_2,n_3}$.
They can be determined demanding the eigenequation to be satisfied at a lattice composed of
fixed points.
To illustrate this, let us restrict for simplicity to the one-dimensional case, rewriting
the ansatz \eqref{eff} as
\be
\label{fa}
f(x) = \sum_{n=1}^{N} c_n f_n(x)\,.
\ee
The eigenequation reads
\be
\label{eigsimp}
\hat{F}_{\alpha \gamma} f(x) = \lambda f(x)\,.
\ee
Let $\{ x_n \}_{n=1}^{N}$ stands for the lattice. For instance, one could consider
Tchebychev's nodes. These are
defined as roots of the Tchebychev polynomial of the first kind of the degree
 $n$ \cite{Grandclement:2007sb}.
On the finite interval $[-1,1]$, they read
\be
\label{cheb0}
x_n = \cos\left(\frac{2n-1}{2N} \pi\right), \quad n = 1,...,N\,.
\ee
This can be extended on $[a,b]$ defining
\be
\label{cheb}
x_n = \frac{a+b}{2} + \frac{b-a}{2} \cos\left(\frac{2n-1}{2N} \pi\right), \quad n = 1,...,N\,.
\ee
Here, it is important that the number of points should match the number of coefficients $c_n$.
At the lattice
Eq. \eqref{eigsimp} can be rewritten as
\be
\label{mateq}
F_{nm}^{\alpha \gamma}\, c^m = \lambda f_{n m}\, c^m,
\ee
where $\vec{c} = (c^n) = \{c_1,...,c_N\}$ is a vector built out of unknown coefficients,
while $(f_{n m})$
and $(F_{nm})$ stand for $N \times N$ matrices defined as
\be
\label{ff}
f_{nm} := f_m(x_n)\,, \quad F_{nm}^{\alpha \gamma} := \hat{F}_{\alpha \gamma} f_m(x)|_{x=x_n}.
\ee
Solving Eq. \eqref{mateq} one guaranties the combination \eqref{fa} satisfies the eigenequation \eqref{eigsimp}
at $x = x_n$, $n = 1,...,N$. Eq. \eqref{mateq} is a generalized eigenequation: having specified matrices $(f_{n m})$
and $(F_{nm}^{\alpha \gamma})$ one obtains unknown coefficients $c_n$ and the eigenvalue $\lambda$ solving algebraic
eigenequation \eqref{mateq}. The coefficients specify approximate solution of the differential equation. The denser
the grid $\{x_n\}$, the better the precision.

Now, we consider 3-dimensional case. The first element of the construction is the functional basis
$f_{n_1 n_2 n_3}(x_1,x_2,x_3)$. It turns out, a convenient choice is the basis
\be
\label{fans}
f_{n_1 n_2 n_3}(x_1,x_2,x_3) = \sin\left( 1 + \frac{\ln|x_1|}{n_1} + \frac{\ln|x_2|}{n_2} + \frac{\ln|x_3|}{n_3} \right).
\ee
We can now search for solutions to the eigenequation \eqref{Feq}. Finding the numerical solution $f_{n_1 n_2 n_3}$ of
Eq. \eqref{Feq}, one finds the solution to the eigenequation \eqref{StEvolEigenEq2}, given by Eq. \eqref{ansa1}.
 In order to do so, consider a  tree-dimensional grid $\{x_n\} = \{x_1^{(n)},x_2^{(n)},x_3^{(n)}\}$. As we are
interested  in covering both positive and negative $x_i \in \mathbb{R}$ in \eqref{fans}, it is reasonable to
allow negative $n_i$ (we exclude $n_i = 0$ because of the form of the right hand side of Eq. \eqref{fans}).
The sum \eqref{fa} becomes $\sum_{n=-N}^{-1} c_n f_n(x) + \sum_{n=1}^N c_n f_n(x)$ . Due to the presence of
logarithmic function in Eq. \eqref{fans}, this choice should respect the fact that terms in Eq. \eqref{fans} become
highly oscillating in the limit $x_i \rightarrow 0$. Hence, it is reasonable to make the grid denser close
to zero. However, this is not the case for the original Tchebychev's nodes \eqref{cheb}. This can be achieved
adopting the following, modified Tchebychev's nodes:

\be
\label{cheb2}
x_n = b_\pm \left( 1 + \cos\left(\frac{2 n-1}{4 N} \pi\right) \right), \quad n = N+1,...,2N\,,
\ee
where $b_\pm$ stands for two real parameters, positive $b_+$ and negative $b_-$. They provide respectively
positive and negative nodes. Clearly, this holds for all three dimensions.  Because of terms
$|x_1 x_2 x_3|^\alpha \exp(-\gamma/2 (|x_1|+|x_2|+|x_3|))$, the function \eqref{ansa1} vanishes close to zero,
$|x_I| \ll 1$, and close to infinity, $|x_I| \gg 1$. Therefore, one gets a good approximation restricting to
a relatively small finite number of nodes.  This justifies the choice \eqref{cheb2}.
Having specified the grid and functional basis, we are ready to find the solutions. Choosing\footnote{ It
turns out that takeing  $|b_-| \neq |b_+|$ significantly improves numerical precision. }
\be
\label{bbN}
b_- = -3, \quad b_+ = 3.5, \quad N=5 \, ,
\ee
and adopting the basis \eqref{fans} with
\be
\label{agam}
\alpha = \frac{1}{2}, \quad \gamma = 1,
\ee
we get approximate solutions with discrete spectrum of positive and negative eigenvalues. Restricting
to negative values and starting with the highest $\lambda$, one finds:
\begin{align}
\nonumber
\lambda \simeq &-2.193, -2.193, -6.470, -6.470, -14.34, -17.71, -27.34, -27.34,
\\[1ex]
\nonumber
&-32.05, -36.26, -39.65, -47.50, -47.50, -62.62, -62.62, -63.83,
\\[1ex]
\nonumber
&-74.97, -102.0, -110.7, -125.1, -125.1, -125.7, -155.9, -249.6,
\\[1ex]
&-249.6, -325.3, -325.3, -358.0, -358.0
\end{align}
For instance, choosing the third  eigenvalue one finds the solution $\psi_{s}$, and the corresponding
numerical error given by Eq. \eqref{localerror}:
\be
\label{spectsym}
\psi_s: \quad \lambda_s = - 6.470, \quad E[\psi_s] \simeq 2.86 \times 10^{-6}.
\ee
Here $\psi_s$ stands for a normalized function found by renormalization of the numerical solution
$ \psi_s \rightarrow \psi_s / \| \psi_s \|^{1/2}$. We have obtained a fairly good precision despite
considering a small grid. More precisely, taking $N=5$ means we adopted ten points per each dimension
(the whole three dimensional lattice is composed of $1000$ points). The rationale for this are the
following. First, the functions \eqref{fans} provides a good basis in the sense that a combination
involving small number of terms results in good approximation to the solution of the eigenequation
(in the sense the numerical error turns out to be small).
This is because of the presence of logarithmic function; something that has already been observed
discussing variational method.  Second, the eigenfunction vanishes
fast for $|x_I| \gg 1$, and so we can restrict our analysis to covering a small, finite
region $x_I \in [b_-,b_+]$.

In addition to the symmetric function \eqref{fans}, one can consider the antisymmetric one,
adopting the basis
\begin{align}
\label{afans}
\nonumber
f_{n_1 n_2 n_3}(x_1,x_2,x_3) = &\left( \textmd{sign}(x_1)+\textmd{sign}(x_2)+\textmd{sign}(x_3) \right) \cdot
\\[1ex]
& \cdot \sin \left( 1 + \frac{\ln|x_1|}{n_1}+\frac{\ln|x_2|}{n_2}+\frac{\ln|x_3|}{n_3}  \right).
\end{align}
Adopting the choices \eqref{bbN}--\eqref{agam} leads to the following spectrum of negative eigenvalues:
\begin{align}
\nonumber
\lambda \simeq &-2.193, -2.193, -6.470, -6.470, -14.34, -17.71, -32.05, -36.26, \
\\[1ex]
\nonumber
&-39.65, -47.50, -47.50, -63.83, -74.97, -96.98, -96.98, -102.0, \
\\[1ex]
&-110.7, -125.7, -155.9, -249.6, -249.6, -325.3, -325.3, -358.0, -358.0.
\end{align}
Choosing, for instance,  the first  eigenvalue, one finds the antisymmetric solution $\psi_a$ and the corresponding
error \eqref{localerror}:
\be
\label{spectasym}
\psi_a: \quad \lambda_a = - 2.193, \quad E[\psi_a] \simeq 2.68 \times 10^{-6}.
\ee
The functions $\psi_s$ and $\psi_s$ are orthogonal and they both were constructed as normalized.

%%%%%%%%%%%%%%%%%%%%%%%%%%%%%%%%%%%%%%%%%%%%%%%%%%%%%%
\subsection{Imposition of the dynamical constraint}
%%%%%%%%%%%%%%%%%%%%%%%%%%%%%%%%%%%%%%%%%%%%%%%%%%%%%%

Eq. \!\eqref{StateEvolutionEq2} is the Schr\"{o}dinger-like equation
corresponding to the classical dynamics defined by
Eqs. \!\eqref{ab3}--\eqref{ab4}. However, the latter is constrained by the condition $H = 0$, with $H$ given by \eqref{b7}.
The Dirac quantization scheme applied in this paper
consists in mapping the classical constraint to the quantum constraint
$\hat{H} = 0$, which according to Eq. \!\eqref{r2} reads:
\begin{equation}
\label{qc1}
\hat{H}\Psi(\phi,x) := \left[ E \left(\frac{\partial}{\partial\phi}\right) + \hat{W} \right] \Psi(\phi,x) = 0\,.
\end{equation}
Therefore, not all solutions to \eqref{StateEvolutionEq2} are physical but only
the ones satisfying \eqref{qc1}. It turns out, however, that the solution to
\eqref{qc1}, in the form \eqref{SeparationVarEvolEq} with $\omega_\lambda$
defined by \eqref{solE1}--\eqref{solE2}, can only be the trivial one $\Psi(x)
= 0$. To address this difficulty, we propose to impose, instead of \eqref{qc1},
the weak form of the Dirac condition:
\begin{equation}
\label{qc2}
\langle \Psi| \hat{H}\Psi \rangle =: \langle \hat{H} \rangle_\Psi = 0\,,
\end{equation}
which has to be satisfied by a given linear combination of the products of eigenfunctions
\begin{equation}\label{qc3}
\Psi(\phi,x) = \SumInt_\lambda
\omega_\lambda(A_\lambda,B_\lambda;\phi)\, \psi_\lambda(x)\,,
\end{equation}
where $\omega_\lambda$ are defined by \eqref{solE1}--\eqref{solE2} (up to arbitrary constants
$A_\lambda$ and $B_\lambda$), and $\psi_\lambda$
is determined numerically via \eqref{Nfun} and \eqref{ansa1}. The symbol
$\SumInt_\lambda$ denotes summation or integration depending on the solutions to
the eigenequations \eqref{StEvolEigenEq}--\eqref{StEvolEigenEq2}.
%Imposing the condition \eqref{qc2}, instead of \eqref{qc1}, means that we propose to satisfy the Dirac condition in a weak form.

Since $\hat{W}$ is a Hermitian operator, we have
$\langle\psi_{\lambda^\prime}|\hat{W}|\psi_{\lambda}\rangle = \lambda\,
\delta(\lambda^\prime, \lambda)$, with $\lambda \in \dR$, so Eq. \!\eqref{qc2}
takes the form
\begin{equation}\label{qc4}
\langle \hat{H} \rangle_\Psi = \SumInt_\lambda
\omega^\ast_{\lambda}(A_\lambda,B_\lambda;\phi)
\left(\lambda - \frac{\kappa}{2}\frac{d^2}{d \phi^2}\right)\, \omega_{\lambda}(A_\lambda,B_\lambda;\phi) = 0\,.
\end{equation}
For the case $\kappa \lambda \neq 1/2$, Eq. \!\eqref{qc4} leads to
\begin{equation}\label{qc5}
\SumInt_\lambda \omega^\ast_{\lambda}(A_\lambda,B_\lambda;\phi)
\big( \omega_{\lambda}(A_\lambda,B_\lambda;\phi) + i \sqrt{2\kappa \lambda - 1}\,
\omega_{\lambda}(A_\lambda,-B_\lambda;\phi) \big) = 0\,,
\end{equation}
whereas for the case $\kappa \lambda = 1/2$ we get
\begin{equation}\label{qc6}
\omega^\ast_\lambda(A_\lambda,B_\lambda;\phi)\, \omega_\lambda(\kappa A_\lambda,\kappa B_\lambda +
 i A_\lambda;\phi) = 0\,.
\end{equation}

In what follows we consider $\kappa = 1$ and $\lambda < 1/2$, in which case $2
\kappa \lambda - 1 < 0$ so that Eq. \!\eqref{solE1} presents an oscillatory
solution.

For $\lambda < 0$, one has
\begin{equation}\label{Con1}
1 + \sqrt{|2\lambda - 1|} > 0 \qquad {\rm and} \qquad 1 - \sqrt{|2\lambda - 1|} < 0\,.
\end{equation}
Eq. \!\eqref{qc5} leads to the condition
\begin{equation}\label{Con2}
\SumInt_\lambda [ (1 - \sqrt{|2\lambda - 1|}) |A_\lambda|^2 + (1 + \sqrt{|2\lambda - 1|}) |B_\lambda|^2 ] = 0\,,
\end{equation}
where $A_\lambda B_\lambda = 0$.
Assuming the orthonormality condition $\langle \psi_{\lambda^\prime}|\psi_{\lambda} \rangle = \delta(\lambda^\prime,\lambda)$, we get
\begin{equation}\label{Con3}
\SumInt_\lambda |\omega_\lambda(A_\lambda,B_\lambda;\phi)|^2 = 1\,.
\end{equation}
Eqs. \!\eqref{Con1}--\eqref{Con3} lead to the condition
\begin{equation}\label{Con4}
\SumInt_{\lambda \in \mathcal{O}_1} |A_\lambda|^2 + \SumInt_{\lambda \in \mathcal{O}_2} |B_\lambda|^2 = 1\,,
\qquad {\rm where} \qquad \mathcal{O}_1 \cap \mathcal{O}_2 = \varnothing\,.
\end{equation}

Let us consider a special solution including only two eigenvalues $\lambda \neq \lambda^\prime$. In such a case
Eqs. \!\eqref{Con2}--\eqref{Con4} give
\begin{equation}\label{Con5}
 \left(1 - \sqrt{|2\lambda - 1|}\right) |A_\lambda|^2 + \left(1 + \sqrt{|2\lambda^\prime - 1|}\right) |B_{\lambda^\prime}|^2 = 0\,,
\end{equation}
and
\begin{equation}\label{Con6}
  |A_\lambda|^2 + |B_{\lambda^\prime}|^2 = 1\,.
\end{equation}
The solution to \eqref{Con5}--\eqref{Con6} reads
\begin{equation}\label{Con7}
  |A_\lambda|^2 = \frac{1 + \sqrt{|2\lambda^\prime - 1|}}{\sqrt{|2\lambda - 1|} + \sqrt{|2\lambda^\prime - 1|}}\,,
  \quad |B_{\lambda'}|^2 = \frac{\sqrt{|2 \lambda -1|}-1}{\sqrt{|2\lambda - 1|} + \sqrt{|2\lambda^\prime - 1|}}\,.
\end{equation}

Therefore, one of the possible solutions to the constraint \eqref{qc2} has the form
\begin{equation}\label{Con8}
\Psi(\phi,x) = \omega_\lambda (A_\lambda,B_\lambda;\phi)\, \psi_\lambda (x) +
\omega_{\lambda^\prime} (A_{\lambda^\prime},B_{\lambda^\prime};\phi)\, \psi_{\lambda^\prime}(x)\,,
\end{equation}
which is defined by the specification of any pair of $\lambda \neq \lambda^\prime$.

%%%%%%%%%%%%%%%%%%%%%%%%%%
\section{Quantum spikes}
%%%%%%%%%%%%%%%%%%%%%%%%%%

The expectation values of our basic observables \eqref{c1} in a state described
by the wavefunction \eqref{qc3} (satisfying the weak Dirac condition \eqref{qc2}) read
\begin{align}\label{Qs1}
\langle\hat{C}_I \rangle (\phi) &= \int_{\dR_+^3} d\nu(x_1,x_2,x_3) \Psi^\star(\phi,x_1,x_2,x_3)
\hat{C}_I \Psi(\phi,x_1,x_2,x_3) \nonumber\\
&= \SumInt_{\lambda_1,\lambda_2} \omega_{\lambda_1}^\star (A_{\lambda_1}, B_{\lambda_1};\phi)\,
\omega_{\lambda_2} (A_{\lambda_2}, B_{\lambda_2};\phi)\,
\langle \psi_{\lambda_1}|\hat{C}_I|\psi_{\lambda_2} \rangle\,,
\end{align}
and
\begin{equation}\label{Qs2}
\langle\hat{D}_I \rangle (\phi) = \SumInt_{\lambda_1,\lambda_2} \omega_{\lambda_1}^\star (A_{\lambda_1},
B_{\lambda_1};\phi)\, \omega_{\lambda_2} (A_{\lambda_2}, B_{\lambda_2};\phi)\,
\langle \psi_{\lambda_1}|\hat{D}_I|\psi_{\lambda_2} \rangle\,.
\end{equation}
The coefficients $A_\lambda$, $B_\lambda$ occurring here can be fixed
by imposing the initial conditions at a certain value of $\phi = \phi_0$ so that for the case $I = 1$
we have\footnote{ Since the classical spike has been derived for the case $I = 1$,
we stick to this case at the quantum level as well. But the system is symmetric with respect to the choice of $I$
so that the same is true for two other cases.}:
\begin{align}\label{Qs3a}
\langle \hat{C}_1 \rangle (\phi_0) =: \tilde{x}_{C_1},
\\[1ex]
\label{Qs3b}
\langle \hat{D}_1 \rangle (\phi_0) =: \tilde{x}_{D_1},
\end{align}
with $\tilde{x}_{C_1},\, \tilde{x}_{D_1} \in \mathbb{R}$.
In principle, $\phi_0$ can be arbitrary but we choose $\phi_0 = 0$ below, as we also did for classical variables in Figs. \!\ref{fig:d1c1phic}-\ref{fig:chis}.

In the next two subsections we apply our numerical results  for  the
wavefunction $\Psi$ to calculate (\ref{Qs1}) and (\ref{Qs2}).

%%%%%%%%%%%%%%%%%%%%%%%%%%%%%%%%%%%%%%%%%%%%%%%%
\subsection{Using the results of the variational method}
%%%%%%%%%%%%%%%%%%%%%%%%%%%%%%%%%%%%%%%%%%%%%%%%%

Let us first consider the solutions \eqref{ans1}--\eqref{nsol1} and \eqref{ans2}--\eqref{nsol2}.  We  calculate
\be
\label{cmat}
(C_1) =  \begin{pmatrix}
0 & c_0 + i \delta c \\
c_0 - i \delta c & 0
\end{pmatrix}\,,
\ee
where $(C_1)_{ij} := \langle \psi_{\lambda_i}|\hat{C}_1| \psi_{\lambda_j}\rangle$, and we get

\be
\label{val1}
c_0 \simeq -0.0174\,, \quad \delta c \simeq -2.21 \cdot 10^{-4}.
\ee

The full wavefunction $\Psi(\phi,x)$ is given by Eq. \eqref{Con8} with $\lambda = \lambda_1$ and $\lambda' = \lambda_2$. Due to the condition
$A_\lambda B_\lambda = 0$ (cf. (\ref{Con2})), we may assume that e.g. $A_{\lambda_2} = B_{\lambda_1} = 0$. Then, $A_{\lambda_1}$ and $B_{\lambda_2}$ remain two independent
complex parameters. Let us parameterize them as
\be
A_{\lambda_1} = |A_{\lambda_1}|\, e^{i \varphi_1}, \quad B_{\lambda_2} = |B_{\lambda_2}|\, e^{i \varphi_2}.
\ee
The absolute values $|A_{\lambda_1}|$, $ |B_{\lambda_2}|$ are fixed by Eqs. \eqref{Con7}:
\begin{equation}\label{Con7b}
|A_{\lambda_1}| = \sqrt{\frac{1 + \sqrt{|2 \lambda_2 - 1|}}{\sqrt{|2 \lambda_1 - 1|} + \sqrt{|2 \lambda_2 - 1|}}}\,,
\quad |B_{\lambda_2}| = \sqrt{\frac{\sqrt{|2 \lambda_1 - 1|} - 1}{\sqrt{|2 \lambda_1 - 1|} + \sqrt{|2 \lambda_2 - 1|}}}\,.
\end{equation}
The considered numerical solutions \eqref{nsol1} and \eqref{nsol2} correspond, respectively, to the values $\lambda_1 \simeq -0.0821$ and $\lambda_2 \simeq -0.0957$.
Since we are interested in the oscillatory case, we choose $\kappa = 1$, so that \eqref{solE1} gives us complex functions
%Plugging $A_{\lambda_2} = B_{\lambda_1} = 0$ and $\kappa = 1$ into Eq. \eqref{solE1} gives
\be
\omega_{\lambda_1}(\phi) = A_{\lambda_1} e^{i \left(\sqrt{|2 \lambda_1-1|} - 1\right) \phi}\,, \quad \omega_{\lambda_2}(\phi) = B_{\lambda_2} e^{-i \left(\sqrt{|2 \lambda_2 - 1|} + 1\right) \phi}\,.
\ee
The final form of the wavefunction reads
\be
\label{psiin}
\Psi(\phi,x) = \omega_{\lambda_1}(\phi)\, \psi_{\lambda_1}(x) + \omega_{\lambda_2}(\phi)\, \psi_{\lambda_2}(x)\,.
\ee

Calculation of the expectation value \eqref{Qs1} for the state \eqref{psiin} leads to the result
\be
\label{Cfinal0}
\langle\hat{C}_1 \rangle (\phi) = \beta \cos(\Delta \varphi + \chi \phi) + \delta \beta \sin(\Delta \varphi + \chi \phi)\,,
\ee
with the following parameters
\begin{align}
\label{dfi}
&\Delta\varphi := \varphi_1 - \varphi_2\,,
\\[1ex]
\label{cfbeta}
&\beta := 2 c_0 |A_{\lambda_1}| |A_{\lambda_2}|\,, \quad \delta\beta = 2 \delta c |A_{\lambda_1}| |A_{\lambda_2}|\,,
\\[1ex]
\label{cfchi}
&\chi := \sqrt{|2 \lambda_1-1|}+\sqrt{|2 \lambda_2-1|}\,.
\end{align}
The values of $\beta$, $\delta\beta$ and $\chi$ for the case of (\ref{val1}) and $\lambda_1,\lambda_2$ mentioned below (\ref{Con7b}) are
\be
\label{val2}
\beta \simeq -0.0065\,, \quad \delta\beta \simeq -8.3 \cdot 10^{-6}, \quad \chi \simeq 2.17\,.
\ee
Meanwhile, the parameter $\Delta\varphi$ can be eliminated from Eq. \eqref{Cfinal0} by imposing on the latter the initial condition \eqref{Qs3a}, which gives
\be
\label{eco}
\beta \cos(\Delta \varphi + \chi \phi_0) + \delta \beta \sin(\Delta \varphi + \chi \phi_0) = \tilde{x}\,,
\ee
where we simplified the notation by taking $\tilde{x} \equiv \tilde{x}_{C_1}$. Eq. \!\eqref{eco} allows  to express $\Delta\varphi$ as a function of $\tilde{x}$.
Assuming that $\phi_0 = 0$, we find two different solutions
$\Delta\varphi^{(\pm)} = \Delta\varphi^{(\pm)}(\tilde{x})$:
\begin{align}
\label{deltaphi}
\Delta\varphi^{(\pm)} &= \textmd{atan2} \left( \frac{\beta \tilde{x} \mp |\delta \beta| \sqrt{\beta^2 + \delta \beta^2
- \tilde{x}^2}}{\beta^2 + \delta \beta^2}, \frac{\delta \beta \tilde{x} \pm |\beta| \textmd{sign}(\delta \beta) \sqrt{\beta^2
+ \delta \beta^2 - \tilde{x}^2}}{\beta^2 + \delta \beta^2} \right) \nonumber\\[1ex]
&+ 2n \pi\,, \quad n \in \mathbb{Z}\,,
\end{align}
where $\textmd{atan2(.,.)}$ stands for two-argument arctangent function.  Substituting \eqref{deltaphi} into Eq. \eqref{Cfinal0}, we ultimately obtain
\be
\label{deltaphi2}
\langle\hat{C}_1 \rangle_{(\pm)}(\phi) = \cos(\chi \phi)\, \tilde{x} \mp  \textmd{sign} ( \delta \beta) \sin(\chi \phi) \sqrt{\beta^2
+ \delta \beta^2 - \tilde{x}^2}\,.
\ee

In order to detect quantum spikes, we now examine $\langle \hat{C}_1 \rangle_{(\pm)}$ for a fixed $\phi$ but as a function of $\tilde{x}$.
The dependence on $\tilde{x}$ is trivial for $\phi = 0$.
For $\phi \neq 0$, the domain is restricted to the interval $\tilde{x} \in [-\sqrt{\beta^2 + \delta\beta^2},\sqrt{\beta^2
+ \delta\beta^2}]$ and the function $\langle \hat{C}_1 \rangle_{(\pm)}(\tilde{x})$ has the non-trivial derivative:
\be
\frac{d \langle \hat{C}_1 \rangle_{(\pm)}(\tilde{x})}{d \tilde{x}} = \cos(\chi \phi) \pm \frac{\textmd{sign} ( \delta \beta) \sin(\chi \phi)\,
\tilde{x}}{\sqrt{\beta^2 + \delta \beta^2 - \tilde{x}^2}}\,.
\ee
In particular, at $\tilde{x} = 0$ we have
\be
\label{derv1}
\frac{d \langle \hat{C}_1 \rangle_{(\pm)}(\tilde{x})}{d \tilde{x}} \Big |_{\tilde{x} = 0} = \cos(\chi \phi)\,,
\ee
which vanishes if
\be
\label{cspike}
\chi \phi = \frac{\pi}{2} + k \pi\,, \quad k \in \mathbb{Z}\,.
\ee
The second derivative of Eq. \eqref{deltaphi2} reads
\be
\label{derv2}
\frac{d^2 \langle \hat{C}_1 \rangle_{(\pm)}(\tilde{x})}{d \tilde{x}^2} = \pm \frac{ (\beta^2+\delta \beta^2)  }{(\beta^2+\delta \beta^2 -
\tilde{x}^2)^{3/2}} \textmd{sign}(\delta \beta) \sin(\chi \phi)\,.
\ee
Eqs. \eqref{derv1}--\eqref{derv2} show that, depending on the values of $\chi \phi$ and $\delta\beta$, $\langle \hat{C}_1 \rangle_{(\pm)}$ reach a local maximum or minimum at $\tilde{x} = 0$. In particular, taking (\ref{val2}) and $k = 0$ in \eqref{cspike}, one finds that this happens for $\phi \simeq 0.72$. We consider such a phenomenon to be the quantum analogue of a classical spike. In general, these quantum spikes occur only at specific moments of time $\phi$, belonging to a periodic discrete set with the period $\Delta\phi = \pi/\chi \simeq 1.45$, determined by Eq. \eqref{cspike}.

\begin{figure}[h]
\includegraphics[width=0.45\textwidth]{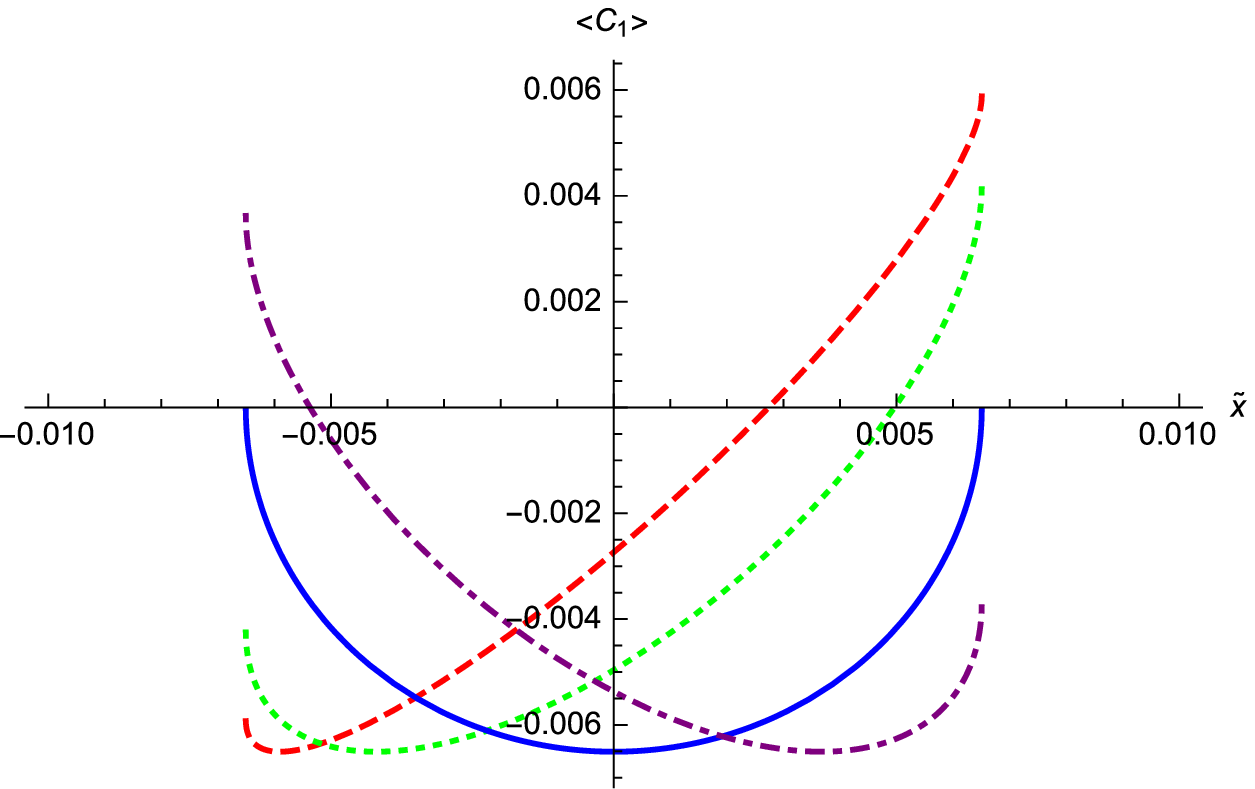}
\hspace{0.05\textwidth}
\includegraphics[width=0.45\textwidth]{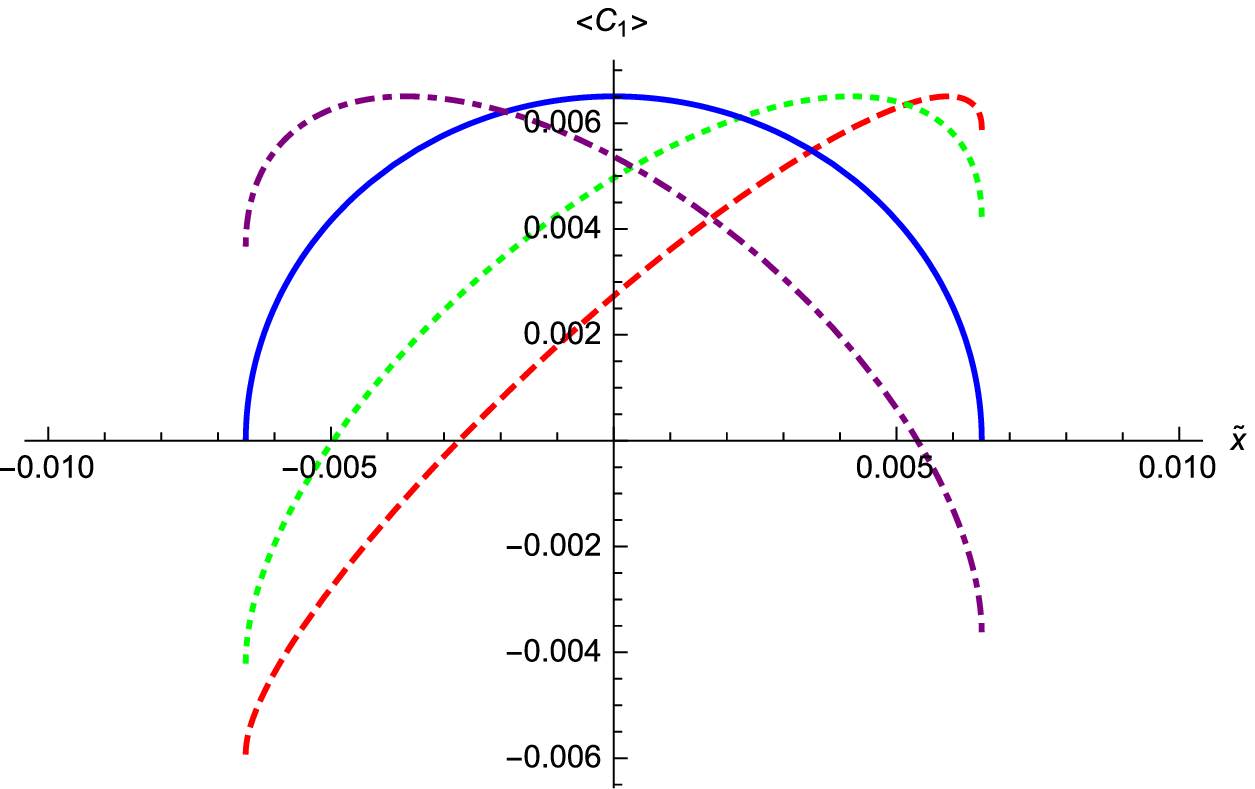}
\caption{\label{fig:Cplot} $\langle \hat{C}_1 \rangle_{(-)}$ (left) and $\langle \hat{C}_1 \rangle_{(+)}$ (right) for evolution
parameters $\phi = 0.2$ (dashed), $\phi = 0.4$ (dotted), $\phi = 0.72$ (solid) and $\phi = 1$ (dash-dotted).}
\end{figure}

The function $\langle \hat{C}_1 \rangle_{(\pm)}(\tilde{x})$ is shown in Fig. \ref{fig:Cplot}. In both cases spikes occurring for
$\phi \simeq 0.72$ are represented by solid lines.

Having calculated $\langle \hat{C}_1 \rangle$, we can repeat the above analysis for $\langle \hat{D}_1 \rangle$. The matrix of elements $(D_1)_{ij} :=
\langle \psi_{\lambda_i}|\hat{D}_1| \psi_{\lambda_j}\rangle$ has the form
\be
\label{dmat}
(D_1) = \begin{pmatrix}
d_1 & 0 \\
0 & d_2
\end{pmatrix}
\ee
where
\be
d_1 \simeq -5.71 \cdot 10^{-6}, \quad d_2 \simeq 7.35 \cdot 10^{-5}.
\ee
Analogously to (\ref{Cfinal0}), we obtain
\begin{align}
\nonumber
\langle \hat{D}_1 \rangle &= \frac{d_1 - d_2 + d_1 \sqrt{|1 - 2 \lambda_2|} + d_2 \sqrt{|1 - 2 \lambda_1|}}{ \sqrt{|1 - 2 \lambda_1|}
+ \sqrt{|1 - 2 \lambda_2|}}
\\[1ex]
&\simeq -2.83 \cdot 10^{-6}.
\end{align}
Here, $\langle \hat{D}_1 \rangle = {\rm const}$ because there are no  off-diagonal components in the matrix \eqref{dmat}. More precisely, the off-diagonal components are non-zero, but are small of order $10^{-20}$.  This result is almost unaffected by change of the order $N$ of the numerical approximation and off-diagonal components are actually becoming smaller with growing\footnote{ For instance, for $N = 8$ one finds them to be equal $\simeq 7.7 \cdot 10^{-20}$, while for $N = 10$ one gets $\simeq 2.2 \cdot 10^{-20}$.} $N$. In conclusion, the numerical results indicate that $\langle \hat{D}_1 \rangle$ does not evolve with time $\phi$.

It is also worth stressing that, at least in the case of restriction to a superposition of two eigenstates \eqref{Con8}, the presence of a spike-like structure, associated with the observable $\hat{C}_1$ is unaffected by the choice of a pair of numerical solutions, i.e. one symmetric and one antisymmetric wavefunction. Adopting different ones modifies the values of coefficients $\beta$, $\delta\beta$, $\chi$ but one can still find the value of $\phi$ corresponding to the quantum spike. In the case of $\hat{C}_1$ the latter is given by Eq. \eqref{cspike}; this is a simple function of two eigenvalues $\lambda_1$, $\lambda_2$. In fact, the crucial requirement for the occurrence of spikes is the presence of non-zero components in the matrix $(C_1)$.

\subsection{Using the results of the spectral method}

One can now perform the analogous analysis for numerical solutions obtained via the spectral method. Taking $\psi_1 = \psi_s$, $\lambda_1 = \lambda_s$
given by Eq. \eqref{spectsym} and $\psi_2 = \psi_a$, $\lambda_2 = \lambda_a$ given by Eq. \eqref{spectasym}, we obtain
\be
\label{cmat2}
(C_1) =  \begin{pmatrix}
0 & c_0 \\
c_0 & 0
\end{pmatrix}\,,
\ee
where $c_0 = -1.28 \cdot 10^{-5}$, while $(D_1)_{ij} = 0,\, \forall i,j$. The matrices $(C_1)$ and $(D_1)$ are defined as before in Eqs. \eqref{cmat} and \eqref{dmat} but in contrast to the former case, the numerical solutions do not contain imaginary terms. For this reason, we have $(D_1) = 0$, as well as $\delta c = 0$. On the other hand, we can achieve the much better precision.

\begin{figure}[h]
\includegraphics[width=0.45\textwidth]{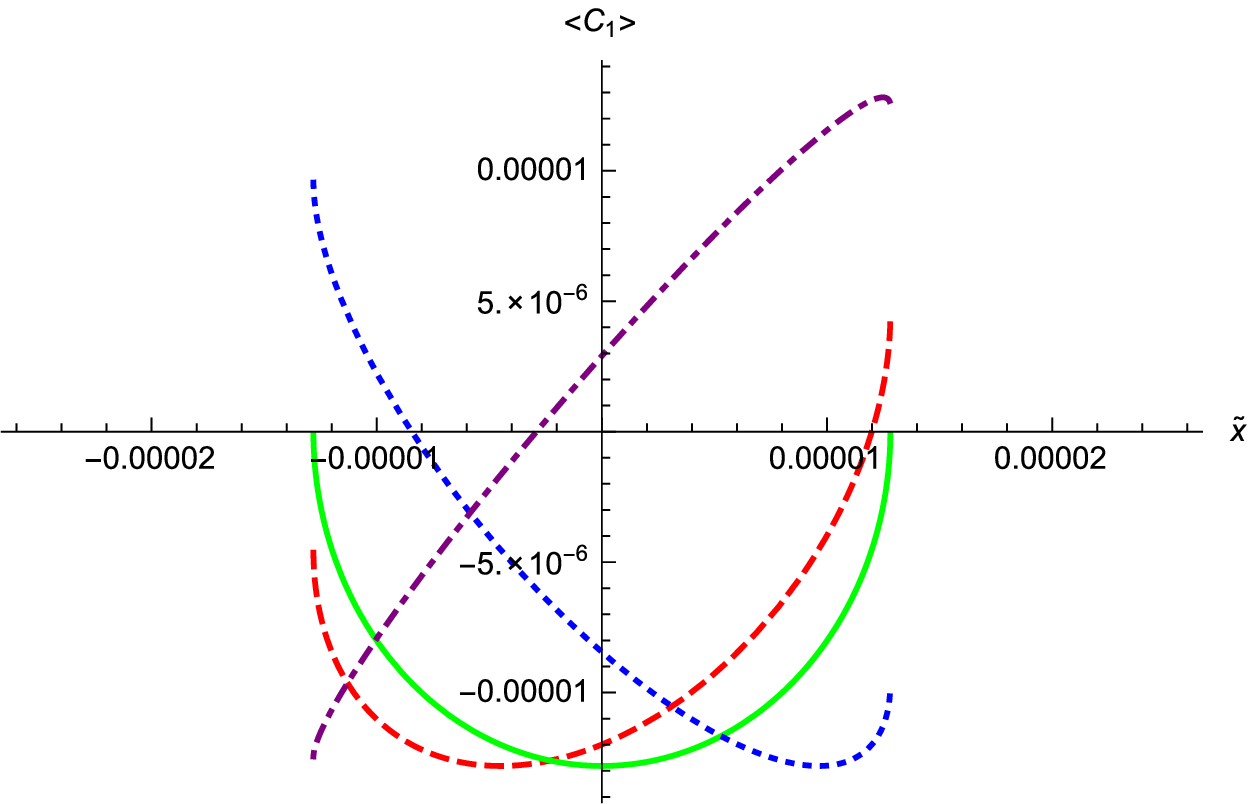}
\hspace{0.05\textwidth}
\includegraphics[width=0.45\textwidth]{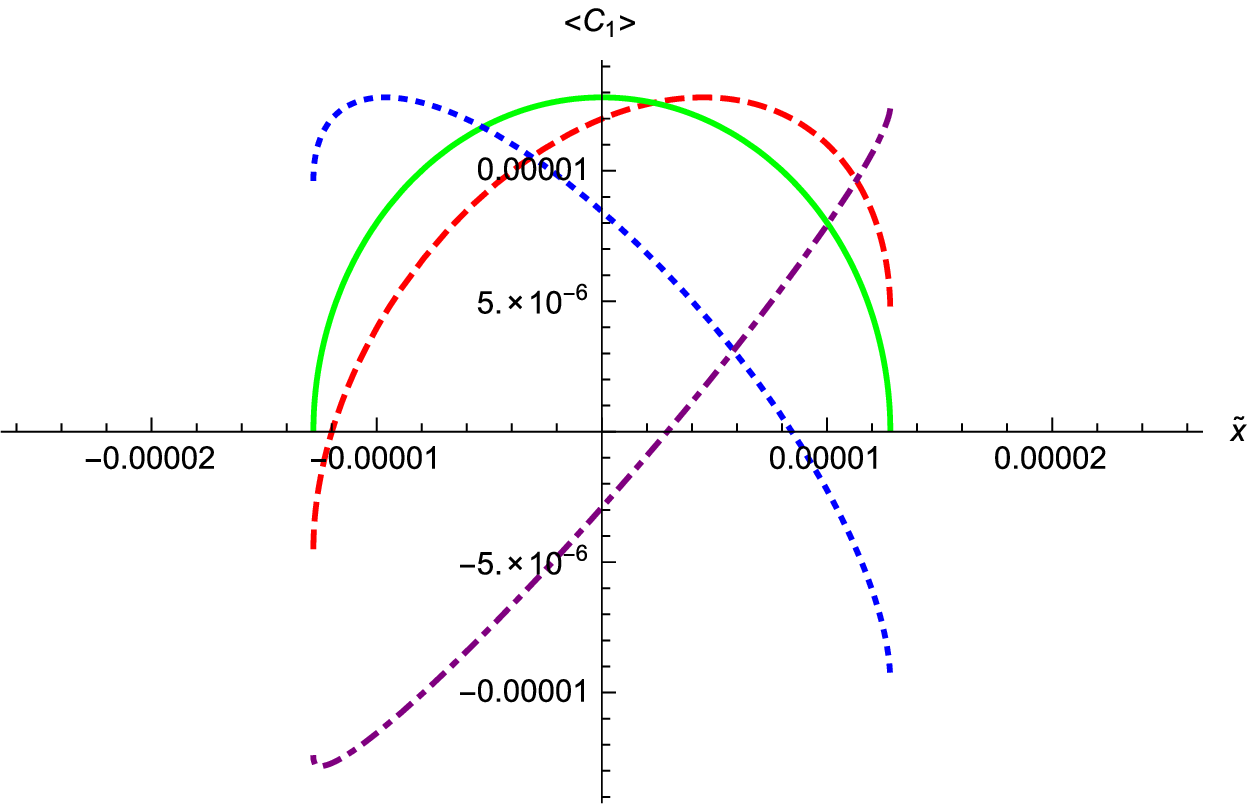}
\caption{\label{fig:Cplot2} $\langle \hat{C}_1 \rangle_{(-)}$ (left) and $\langle \hat{C}_1 \rangle_{(+)}$ (right) for evolution
parameters $\phi=0.2$ (dashed), $\phi=0.26$ (solid), $\phi=0.4$  (dotted), $\phi=1$ (dash-dotted).}
\end{figure}

Following the same steps as described in the previous subsection, one can again express the expectation value $\langle \hat{C}_1 \rangle$ as a simple function of $\phi$ (cf. (\ref{Cfinal0})):
\be
\label{Cfinal0s}
\langle \hat{C}_1 \rangle (\phi) = \beta \cos(\Delta \varphi + \chi \phi)\,,
\ee
where $\Delta\varphi$, $\beta$ and $\chi$ are given by Eqs. \eqref{dfi}--\eqref{cfchi}.  The numerical
values of the latter two constants for the considered case of \eqref{spectsym} and \eqref{spectasym} are
\be
\beta \simeq 1.28 \cdot 10^{-5}, \quad \chi \simeq 6.05\,.
\ee
Similarly to the Eq. \eqref{eco}, we eliminate the parameter $\Delta\varphi$ by imposing the boundary
condition \eqref{Qs3a}, which now becomes
\be
\label{tosols}
\beta \cos(\Delta \varphi + \chi \phi) = \tilde{x}\,.
\ee
 Solving this equation for $\Delta\varphi$, one finds two solutions (as in \eqref{deltaphi} before)
\be
\label{labs}
\Delta\varphi^{(\pm)} = \pm \arccos\left( \frac{\tilde{x}}{\beta} \right) + 2n \pi\,, \quad n \in \mathbb{Z}\,.
\ee
Substitution of $\Delta\varphi = \Delta\varphi^{(\pm)}$ into Eq. \eqref{Cfinal0s} finally gives
\be
\langle\hat{C}_1 \rangle_\pm(\phi) = \tilde{x} \cos( \chi \phi ) \mp \beta
\sqrt{1 - \frac{\tilde{x}^2}{\beta^2}}\, \sin(\chi \phi)\,.
\ee
The two obtained solutions are depicted in Fig. \ref{fig:Cplot2}. Quantum spikes (represented by solid lines) occur in both cases at time $\phi \simeq 0.26$ and are periodic in $\phi$, with the period $\Delta\phi = \pi/\chi \simeq 0.52$.
The existence of spikes is ensured by the presence of non-zero elements in the matrix $(C_1)$.
Starting with a different pair of symmetric and antisymmetric functions $\psi_s$ and $\psi_a$,
corresponding to different eigenvalues $\lambda_1$ and $\lambda_2$, one also expects to find spikes
(unless $(C_1)_{ij} = 0,\, \forall i,j$).

\section{Conclusions}

In this paper, we have attempted to uncover the existence of quantum strange
spikes (in short, quantum spikes), i.e. certain distinctive features in the
quantum evolution of the considered gravitational system. Such features are
expected to be analogues of steep structures, called by us strange spikes, that
arise in the corresponding classical evolution.  Fig.~\!\ref{fig:d1c1phic} shows
that the classical spikes presented in Fig.~\!\ref{fig:d1c1phi} (and previously
in the paper \cite{Czuchry:2016rlo}) are not apparent effects of the projection
of a 3D plot into a 2D plot, but real structures. Furthermore, it seems that an
extremum (maximum, minimum) or an inflection point, occurring locally at the
classical level, may turn into a similar structure at the quantum level.
Quantization does not need to suppress the classical spikes as it was
preliminarily concluded in \cite{Czuchry:2016rlo}.

Let us discuss the latter claim in more detail. On the basis of numerical
results, we conjecture that a quantum spike is an extremum in the evolution of
the expectation value of a given quantum observable.  The inflection-type
classical spike $C_1(\tilde x,\phi)$ presented in Figs.~\!\ref{fig:d1c1phic} and
\ref{fig:d1c1phi} becomes the extremum-type quantum spike presented in
Figs.~\!\ref{fig:Cplot} or \ref{fig:Cplot2}. Another difference is that our
classical spike is a monotonic function of time, whereas the corresponding
candidate for a quantum spike is periodic. To be specific, we have restricted
our analyses to just one quantum observable $\hat{C}_1$ (the results for
$\hat{D}_1$ have insufficient accuracy).

Compared to the classical spikes, quantum spikes differ in two ways. Namely,
they are rather mild and periodic in time. At the quantum level, classical
structures become smoother and of slightly different type, being specific only
to discrete moments in time. Nevertheless, it can be conjectured that classical
spikes survive quantization in this sense.

Constructing solutions of the quantum evolution, we have applied the variational
and spectral methods, which are quite different. These methods are not only
different conceptually but also use different bases of auxiliary
functions. Still, the obtained quantum spikes are quite similar. In particular,
they are periodic in time.  The two completely dissimilar methods lead to
similar structures, which may serve as the robustness test of our results.

For simplicity, we have identified quantum spikes by making use of only two
classes (obtained by two different numerical methods) of solutions to the
quantum dynamics.  Many other classes of solutions are possible and hence other
types of quantum spikes could exist. Some of them might look similar to
classical spikes and be monotonic functions of time. However, increasing the
number of solutions in a wave packet makes the construction much more
technically involved. This is the reason we have restricted ourselves to a pair
of numerical solutions, but it was sufficient to test the method.

Another issue that we need to stress is our application of a classical massless
scalar field in the role of a clock at both the classical and quantum
levels. The scalar field is not coupled to the gravitational degrees of freedom
in the Hamiltonian \eqref{b7} and hence such a treatment is justified. In fact,
this allows us to avoid the inconsistency that is commonly ignored in
literature: the situation when time is a parameter in the classical theory but a
quantized variable in the quantum theory.  In both cases it should be just the
same evolution parameter.  In this paper we did not wish to consider the
quantization of time.

The implementation of the dynamical constraint at the quantum level has been
performed in the weak sense. Such a way of imposing constraints is practised in
other branches of quantum physics and quantum chemistry, especially in
variational methods (see, e.g. \cite{Chong,Smilga,Mang} and references therein).

A study of the corresponding issues in the case of inhomogeneous spacetimes
would be highly interesting since they naturally favour structure
formation. Different to our strange spikes, the spikes found in such spacetimes
(see \cite{sp1,sp2,sp3,sp4,sp5,sp6,sp7,sp8,sp9} and references therein) have
never been quantized. Let us also stress that we do not investigate the possible
relation between the latter ``inhomogeneous'' spikes and our ``homogeneous''
ones, as it is beyond the scope of the present paper.

When we think about the real spikes, which might occur in the early observed
universe, we rather think in terms of possible structures in spacetime. In
contrast, the spikes that we study in this paper arise in the phase space of the
Hamiltonian framework.  The path from the dynamics in phase space to the
dynamics in spacetime is complicated due to the Hamiltonian constraint. Apart
from this, the truncation of the full system to the homogeneous sector
considered in \cite{Ashtekar:2011ck}, which underlies our paper, introduces
additional complexity. Thus, the interpretation of our spikes in terms of the
spikes in spacetime is rather difficult. These difficulties are enhanced by the
procedure of quantization. We postpone the examination of quantum spikes that
are described directly in spacetime to our future work on the quantization of
spikes known in the context of Gowdy space \cite{sp1}.

Our paper is about the possible existence of quantum spikes. The theoretical
framework has been established but much more effort is necessary to prove that
quantum spikes are a generic feature of the quantum gravitational systems. In
particular, the preliminary results presented here could be extended by the
further examination of the eigenequation problem \eqref{StEvolEigenEq2}. New
classes of its solutions could lead to new types of quantum spikes. This
extension of our research definitely requires making use of sophisticated
analytical and numerical tools so that is far from being complete. More activity
in this direction is needed.

%%%%%%%%%%%%%%%%%%%%%%%%%%%%%%%%%%%%%%%%%%%%%%%%%%%%%%%%%%%%%%%%%%%%%%%%%%%%%
%%%%%%%%%%%%%%%%%%%%%%%%%%%%%%%%%%%%%%%

\acknowledgments We would like to thank Vladimir Belinski, Piotr Garbaczewski,
David Garfinkle, Woei Chet Lim, David Sloan, and Claes Uggla for helpful
discussions.

\appendix

\section{Numerical solutions}
\label{num}

All coefficients have been found using Mathematica computer software.

The coefficients $c_{n_1 n_2 n_3}$ of $\psi_S $ corresponding to \eqref{ans1} read:
\begin{align}
\nonumber
&c_{0\, 0\, 0} = 0.0000971398, c_{0\, 0\, 1} = -0.0000668377,
c_{0\, 0\, 2} = 0.0000576278,
\\[1ex]
\nonumber
&c_{0\, 1\, 1} = 0.000145685, c_{0\, 0\, 3} = -0.0000408203,
c_{0\, 1\, 2} = -0.000175153,
\\[1ex]
\nonumber
&c_{1\, 1\, 1} = -0.000093447, c_{0\, 0\, 4} = 5.14783 \times 10^{-6},
c_{0\, 1\, 3} = 0.000120379,
\\[1ex]
\nonumber
&c_{0\, 2\, 2} = 3.38894 \times 10^{-6}, c_{1\, 1\, 2} = 0.000204403,
c_{0\, 0\, 5} = 6.14349 \times 10^{-6},
\\[1ex]
\nonumber
&c_{0\, 1\, 4} = -0.0000357561, c_{0\, 2\, 3} = -0.0000709548,
c_{1\, 1\, 3} = -0.000189175,
\\[1ex]
\nonumber
&c_{1\, 2\, 2} = 0.0000118927, c_{0\, 0\, 6} = 0.0000130402,
c_{0\, 1\, 5} = -7.68871 \times 10^{-6},
\\[1ex]
\nonumber
&c_{0\, 2\, 4} = -0.000037775, c_{0\, 3\, 3} = -0.0000601953,
c_{1\, 1\, 4} = 0.0000173017,
\\[1ex]
\nonumber
&c_{1\, 2\, 3} = -0.00011458, c_{2\, 2\, 2} = 0.0000716101,
c_{0\, 0\, 7} = 9.86615 \times 10^{-6},
\\[1ex]
\nonumber
&c_{0\, 1\, 6} = 0.0000427615, c_{0\, 2\, 5} = -0.0000538152,
c_{0\, 3\, 4} = 0.0000404286,
\\[1ex]
\nonumber
&c_{1\, 1\, 5} = 0.000035388, c_{1\, 2\, 4} = 0.000202186,
c_{1\, 3\, 3} = 0.0000730395,
\\[1ex]
\nonumber
&c_{2\, 2\, 3} = 0.00011985, c_{0\, 0\, 8} = 3.4044 \times 10^{-6},
c_{0\, 1\, 7} = 0.0000249608,
\\[1ex]
\nonumber
&c_{0\, 2\, 6} = -0.0000690364, c_{0\, 3\, 5} = -0.0000391754,
c_{0\, 4\, 4} = -0.0000453004,
\\[1ex]
\nonumber
&c_{1\, 1\, 6} = -0.000104671, c_{1\, 2\, 5} = -0.0000730649,
c_{1\, 3\, 4} = -0.0000535717,
\\[1ex]
\nonumber
&c_{2\, 2\, 4} = 0.0000549831, c_{2\, 3\, 3} = -0.0000756014,
c_{0\, 0\, 9} = 5.96683 \times 10^{-7},
\\[1ex]
\nonumber
&c_{0\, 1\, 8} = 5.7956 \times 10^{-6}, c_{0\, 2\, 7} = -0.0000289886,
c_{0\, 3\, 6} = -0.0000465147,
\\[1ex]
\nonumber
&c_{0\, 4\, 5} = -0.0000499186, c_{1\, 1\, 7} = -0.0000613235,
c_{1\, 2\, 6} = -0.000175011,
\\[1ex]
\nonumber
&c_{1\, 3\, 5} = -0.000150828, c_{1\, 4\, 4} = -0.0000280041,
c_{2\, 2\, 5} = -0.000230914,
\\[1ex]
\nonumber
&c_{2\, 3\, 4} = 0.0000408805, c_{3\, 3\, 3} = 0.0000680184,
c_{0\, 0\, 10} = 4.36818 \times 10^{-8},
\\[1ex]
\nonumber
&c_{0\, 1\, 9} = 5.0905 \times 10^{-7}, c_{0\, 2\, 8} = -4.19215 \times 10^{-6},
c_{0\, 3\, 7} = -0.0000128081,
\\[1ex]
\nonumber
&c_{0\, 4\, 6} = -0.0000245689, c_{0\, 5\, 5} = -0.0000213292,
c_{1\, 1\, 8} = -9.56085 \times 10^{-6},
\\[1ex]
\nonumber
&c_{1\, 2\, 7} = -0.0000502355, c_{1\, 3\, 6} = -0.000111775,
c_{1\, 4\, 5} = -0.0000995701,
\\[1ex]
&c_{2\, 2\, 6} = -0.000193417, c_{2\, 3\, 5} = -0.000350751,
c_{2\, 4\, 4} = -0.000167525.
\end{align}

Similarly, the coefficients $c_{n_1 n_2 n_3}$ of $\psi_A $ corresponding to \eqref{nsol2} read:

\begin{align}
\nonumber
&c_{0\, 0\, 0} = 0.0000154524, c_{0\, 0\, 1} = -7.25196 \times 10^{-7},
c_{0\, 0\, 2} = -6.31044 \times 10^{-6},
\\[1ex]
\nonumber
&c_{0\, 1\, 1} = -1.84189 \times 10^{-6}, c_{0\, 0\, 3} = 7.81442 \times 10^{-6},
c_{0\, 1\, 2} = 4.7519 \times 10^{-6},
\\[1ex]
\nonumber
&c_{1\, 1\, 1} = -6.32555 \times 10^{-6}, c_{0\, 0\, 4} = -6.01093 \times 10^{-6},
c_{0\, 1\, 3} = -0.000016969,
\\[1ex]
\nonumber
&c_{0\, 2\, 2} = -6.35142 \times 10^{-6}, c_{1\, 1\, 2} = 8.55031 \times 10^{-6},
c_{0\, 0\, 5} = -7.06458 \times 10^{-7},
\\[1ex]
\nonumber
&c_{0\, 1\, 4} = 0.000023153, c_{0\, 2\, 3} = -2.43363 \times 10^{-6},
c_{1\, 1\, 3} = -0.0000170932,
\\[1ex]
\nonumber
&c_{1\, 2\, 2} = -0.0000151945, c_{0\, 0\, 6} = -5.85877 \times 10^{-6},
c_{0\, 1\, 5} = -6.02818 \times 10^{-6},
\\[1ex]
\nonumber
&c_{0\, 2\, 4} = 2.59351 \times 10^{-6}, c_{0\, 3\, 3} = 7.14048 \times 10^{-6},
c_{1\, 1\, 4} = 0.0000165033,
\\[1ex]
\nonumber
&c_{1\, 2\, 3} = 0.0000425176, c_{2\, 2\, 2} = 5.83009 \times 10^{-6},
c_{0\, 0\, 7} = -3.85805 \times 10^{-6},
\\[1ex]
\nonumber
&c_{0\, 1\, 6} = -3.95188 \times 10^{-7}, c_{0\, 2\, 5} = -0.0000125786,
c_{0\, 3\, 4} = -0.0000220554,
\\[1ex]
\nonumber
&c_{1\, 1\, 5} = -0.0000163389, c_{1\, 2\, 4} = -0.0000196522,
c_{1\, 3\, 3} = -0.0000324696,
\\[1ex]
\nonumber
&c_{2\, 2\, 3} = -0.0000273951, c_{0\, 0\, 8} = -8.02081 \times 10^{-7},
c_{0\, 1\, 7} = 0.0000107185,
\\[1ex]
\nonumber
&c_{0\, 2\, 6} = 1.27489 \times 10^{-6}, c_{0\, 3\, 5} = 9.31717 \times 10^{-6},
c_{0\, 4\, 4} = 5.99047 \times 10^{-6},
\\[1ex]
\nonumber
&c_{1\, 1\, 6} = -5.39372 \times 10^{-7}, c_{1\, 2\, 5} = -1.92239 \times 10^{-6},
c_{1\, 3\, 4} = 0.0000483597,
\\[1ex]
\nonumber
&c_{2\, 2\, 4} = 0.0000255206, c_{2\, 3\, 3} = 0.0000448492,
c_{0\, 0\, 9} = -3.69269 \times 10^{-8},
\\[1ex]
\nonumber
&c_{0\, 1\, 8} = 4.74694 \times 10^{-6}, c_{0\, 2\, 7} = 5.24472 \times 10^{-6},
c_{0\, 3\, 6} = 7.44385 \times 10^{-6},
\\[1ex]
\nonumber
&c_{0\, 4\, 5} = 6.93398 \times 10^{-6}, c_{1\, 1\, 7} = 6.12731 \times 10^{-6},
c_{1\, 2\, 6} = 0.0000300494,
\\[1ex]
\nonumber
&c_{1\, 3\, 5} = -0.0000468598, c_{1\, 4\, 4} = -0.0000161307,
c_{2\, 2\, 5} = -0.0000284781,
\\[1ex]
\nonumber
&c_{2\, 3\, 4} = -0.000107147, c_{3\, 3\, 3} = -0.0000231205,
c_{0\, 0\, 10} = 3.87598 \times 10^{-9},
\\[1ex]
\nonumber
&c_{0\, 1\, 9} = 6.21427 \times 10^{-7}, c_{0\, 2\, 8} = 1.58561 \times 10^{-6},
c_{0\, 3\, 7} = 1.98172 \times 10^{-6},
\\[1ex]
\nonumber
&c_{0\, 4\, 6} = 0.0000147037, c_{0\, 5\, 5} = -3.09361 \times 10^{-6},
c_{1\, 1\, 8} = 2.09284 \times 10^{-6},
\\[1ex]
\nonumber
&c_{1\, 2\, 7} = 0.0000119165, c_{1\, 3\, 6} = 0.0000362559,
c_{1\, 4\, 5} = 8.76622 \times 10^{-6},
\\[1ex]
&c_{2\, 2\, 6} = 0.0000623312, c_{2\, 3\, 5} = 0.000147871,
c_{2\, 4\, 4} = 0.000069584.
\end{align}

%%%%%%%%%%%%%%%%%%%%%%%%%%%%%%%%%%%%%%%%%%%%%%%%%%%%%%%%%%%%%%%%%%%%%%%%%%

\end{document}